\documentclass[prb,preprint]{revtex4-1}

\usepackage{CJK}
\usepackage{graphicx}

\begin{document}
\begin{CJK*}{UTF8}{gbsn}
\title{Symmetry-dependent electron-electron interaction in coherent tunnel junctions resolved by zero bias anomaly measurements}

\author{Liang Liu (刘亮)}
\author{Jiasen Niu (牛佳森)}
\author{Li Xiang (向黎)}
\author{ Jian Wei (危健)}\email{weijian6791@pku.edu.cn}
\affiliation{International Center for Quantum Materials, School of Physics, Peking University, Beijing 100871, China}
\affiliation{Collaborative Innovation Center of Quantum Matter, Beijing, China}
\author{D.-L. Li}
\author{J.-F. Feng}\email{jiafengfeng@iphy.ac.cn}
\author{X.-F. Han}
\affiliation{Beijing National Laboratory of Condensed Matter Physics, Institute of Physics, Chinese Academy of Sciences, Beijing 100190, China}
\author{X.-G. Zhang}
\affiliation{Center for Nanophase Materials Sciences, and Computer Science and Mathematics Division, Oak Ridge National Laboratory, Oak Ridge, Tennessee 37831-6493, USA}
\affiliation{Department of Physics and Quantum Theory Project, University of Florida, Gainesville 32611, USA}
\author{J. M. D. Coey}
\affiliation{CRANN and School of Physics, Trinity College, Dublin 2, Ireland}

\date{\today}

\begin{abstract}
We provide conclusive experimental evidence that zero bias anomaly in the differential resistance of magnetic tunnel junctions (MTJs) is due to electron-electron interaction (EEI), clarifying a long standing issue. Magnon effect that caused confusion is now excluded by measuring at low temperatures down to 0.2 K and with reduced AC measurement voltages down to 0.06 mV. The normalized change of conductance is proportional to $\ln{(eV/k_{B}T)}$, consistent with the Altshuler-Aronov theory of tunneling that describes the reduction of density of states due to EEI, but inconsistent with magnetic impurity scattering. The slope of the $\ln{(eV/k_{B}T)}$ dependence is symmetry dependent: the slopes for P and AP states are different for coherent tunnel junctions with symmetry filtering, while  nearly the same for those without symmetry filtering (amorphous barriers). This observation may be helpful for verifying symmetry preserved filtering in search of new coherent tunneling junctions, and for probing and separating electron Bloch states of different symmetries in other correlated systems.

\end{abstract}
\maketitle
\end{CJK*}

\section{Introduction}

The prediction of symmetry-preserved tunneling  and the resulting giant tunneling magnetoresistance (TMR) through crystalline MgO barrier in magnetic tunnel junctions (MTJs)~\cite{Butler2001prb,Mathon2001prb} and its subsequent discovery~\cite{Parkin2004nmat,Yuasa2004nmat} represent a major triumph of mean-field theory of electronic structure.~\cite{Zhang2004prb,Yuasa2007jpd} Nowadays  MTJ based on symmetry-preserved tunneling is used for read heads of hard disk drive and is also proposed for future data storage. The key concept is that the magnetoresistance (MR) ratio is determined not only by the spin polarization (SP) of the ferromagnetic electrodes, but also by the matching of symmetries of the Bloch states tunneling through the barrier. This explains why previously the SP obtained from the band calculations did not agree with the estimated SP from tunneling experimental results when interpreted using the traditional Julliere's model,~\cite{Julliere1975} where $MR=2P_{1}P_{2}/(1-P_{1}P_{2})$ is only determined by the SP of electrodes. The inadequacy of Julliere's model for spin-dependent tunneling was first pointed out by MacLaren \textit{et al} in 1997\cite{MacLaren1997prb} and later detailed in the reviews.~\cite{Yuasa2007jpd,Miao2011rpp} 

In recent years, half metals which have full SP ($P=1$) according to band calculations  has draw a lot attention, especially the Heusler alloys.~\cite{Bai2012spin} Theoretically with $P=1$ the MR ratio can be infinity. However, it is found even for the half metal full-Heusler $Co_{2}FeAl_{0.5}Si_{0.5}$ alloys, MR ratio is not ideal when quality of the tunneling barrier or the matching between the tunneling barrier and the electrodes is not perfect,~\cite{Sukegawa2009prb} while excellent MR ratio is achieved when the matching is good.~\cite{Wang2009apl} Thus coherent tunneling is an indispensable factor to get high MR ratio.~\cite{Bai2012spin} While the SP can be directly observed by in situ spin-resovled photoemission spectroscopy,~\cite{Jourdan2014ncomms} there seems no reliable way to verify the coherent tunneling feature.  The oscillation of TMR as a function of MgO thickness was found in cases when the MR ratio is large, and thought to be related to coherent tunneling,~\cite{Wang2010prb_Coherent} but theoretical investigation indicates that it requires the presence of nonspecular scattering inside the barrier that tends to diminish symmetry filtering.~\cite{Zhang2008prb} Here we show by conventional resistance measurement near zero bias, it might be possible to verify the symmetry selective filtering property. 

Resistance measurement near zero bias is not new. In fact the so-called zero bias anomaly has been found for many systems since 1960's.~\cite{Rowell1966prl,Appelbaum1967pr,Appelbaum1966prl,Shen1968pr,Gloos2009ltp} The decrease of MR with increasing bias was observed and has been intensively investigated as to preserve the MR ratio to higher bias is critical for practical use. The origin of the decrease of MR is usually  explained in terms of noninteracting electron picture.
For example, the decrease of MR with increasing bias over the range of a few hundred mV for both AlO$_{x}$-based~\cite{Moodera1995prl,Gallagher1997jap} and MgO-based MTJs~\cite{Matsumoto2005ssc,Ikegawa2007jap} was interpreted as due to inelastic spin non-conserving magnon emission~\cite{Zhang1997prl} or elastic tunneling with variations of tunneling transmission and density of states.~\cite{Cabrera2002apl} 
Electron-electron interaction (EEI) has largely been  neglected in the study of MTJs.

In addition to the slow decrease of MR over the range of a few hundred mV,  for many MTJ samples, both P (when the two electrodes are magnetically aligned) and AP (when the electrodes are magnetically antiparallel) states display an additional resistance cusp within a few tens of mV near zero bias that becomes sharper at lower temperatures. The mechanism behind this additional resistance peak, which we specifically refer as zero bias anomaly (ZBA) here, is a long standing puzzle.~\cite{Bernos2010prb,Nowak2000tsf} Related to ZBA is the `ZB' peak in the second derivative of the current-voltage curves, which is also called inelastic electron tunneling spectroscopy (IETS) and is frequently used to investigate inelastic tunneling processes. In IETS of MTJs,~\cite{Han2001apl,Bang2009jap,Ma2011prb,Drewello2012jap,Teixeira2012apl} the `ZB' peak has been ascribed to various mechanisms, including magnon excitation,~\cite{Miao2006jap}, magnetic impurity scattering,~\cite{Drewello2009prb} and combination of magnon scattering and EEI.~\cite{Bernos2010prb} Although some earlier works~\cite{Zhang1997prl,Moodera1998prl} did not distinguish between ZBA and magnon induced reduction of TMR over a much wider bias range, it was later shown theoretically that interface magnon scattering does not yield sharp peaks at all,~\cite{Wei2010prb} contradicting the interpretation that the ZBA and the sharp peak in the IETS near zero bias are due to magnon excitation. More importantly, there have been suggestions that the ZBA may be due to EEI~\cite{Bernos2010prb,Nowak2000tsf} but conclusive evidence is lacking.

EEI is just the Coulomb interaction between electrons. As proposed by Altshuler and Aronov as early as 1979,~\cite{Altshuler1979,Altshuler1980,Altshuler1985} and later quoted in the textbook by Abrikosov,~\cite{Abrikosov1988} the exchange interaction between electrons can cause quantum corrections to the conductivity as well as density of states. The correction due to the interference of the states has a characteristic  time scale depending on the energy $\epsilon$ involved, and depends on the probability of two particles meeting at the same point, thus on the dimensionality of the systems. Similar arguments also apply for decoherence etc and the dimension related integral $\sim\sqrt{\epsilon}$ in 3D and $\sim\ln{(\epsilon)}$ in 2D.
EEI induced reduction of density of states near the Fermi energy has been demonstrated to be consistent with the observed ZBA in tunneling measurements in the 80's.~\cite{Gershenzon1986jetp} and more recently it is also been considered for cuprates and manganites.~\cite{Abrikosov2000prb,Mazur2007prb} 
However, study of EEI effect concentrated mostly outside of MTJs. The lack of the control over the tunneling wave function symmetry in conventional tunnel junctions means that there has not been any study on whether the EEI effect depends on such symmetry. Symmetry-preserved tunneling in MTJs provides a unique
platform to answer this question and the nearly bias independent resistance in the parallel state is a big advantage for observing ZBA. Yet one must still first settle the debate of whether the ZBA is caused by magnons, magnetic impurities, or EEI.

In this work we first show that the ZBA and the `ZB' peak in IETS are indeed caused by EEI,  then demonstrate that the EEI effect depends sensitively on symmetry filtering in the barrier layer. By performing measurements at subkelvin temperatures and with very small AC voltages, we show that the ZB peaks move to as low as 0.1 mV where contribution from magnon excitation can be excluded and extra broadening due to temperature, modulation, and especially extrinsic noises is minimized. The ZBA is shown to follow the $\ln{(eV/k_{B}T)}$ dependence with decreasing temperature and bias, consistent with the Altshuler-Aronov theory of tunneling in the presence of EEI but inconsistent with the magnetic impurity scattering model.  The slope of the $\ln{(eV/k_{B}T)}$ dependence is different between the P and AP states, suggesting that Bloch states with different symmetries affect the EEI differently. Angular dependence of the slope is also consistent with the EEI mechanism and a separate effect for each incident Bloch wave.

\section{Methods}

\begin{figure}
\includegraphics[width=9cm]{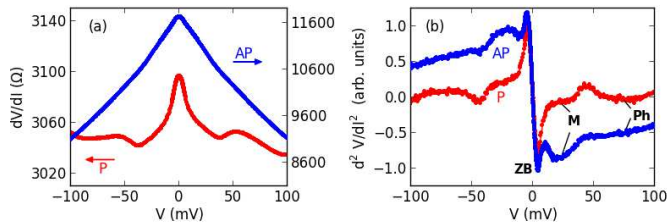}
\caption  {\small (color online) First (a) and second (b) derivative at 3.6 K in the P (red line, 500 G) and  AP (blue line, -300 G) states.  `ZB', `M', and `Ph' peaks in IETS are clearly visible. The maximum  modulation voltage is about 4 mV. } 
\label{fig_dVdI}
\end{figure}

MgO and AlO$_{x}$ MTJ stacks with the main structures of Ir$_{22}$Mn$_{78}$ (10), Co$_{90}$Fe$_{10}$ (2.5), Ru (0.9), Co$_{40}$Fe$_{40}$B$_{20}$ (3), MgO (2.5), Co$_{40}$Fe$_{40}$B$_{20}$ (3) and Co$_{40}$Fe$_{40}$B$_{20}$ (4), AlO$_{x}$ (1), Co$_{40}$Fe$_{40}$B$_{20}$ (4), Ir$_{22}$Mn$_{78}$ (12)(numbers in parentheses indicate nominal thickness in nanometers) were grown with a high vacuum Shamrock cluster deposition tool and an ULVAC magnetron sputtering system, respectively. Then the stacks were patterned into junctions with the rectangular shape of $5\times 10$ $\mu m^{2}$ using ultraviolet (UV) lithography and Ar ion beam etching. After this step, MgO MTJs were annealed in an in-plane field of 8000 Oe at 350 $ ^{\circ}$C for half an hour to define the exchange bias of the antiferromagnetic IrMn layer and crystallize both the bottom and top CoFeB electrodes. The AlOx MTJs were measured in the as-grown state.
Table I summarizes properties of the measured devices, and all figures plotted are for one device M3.

\begin{table}
\caption{Resistance and MR ratio ((R$_{AP}$-R$_{P}$)/R$_{P}$) measured at 4 K for both MgO (M2-M5) and AlO$_{x}$-based (A1-A2) junctions. S$_{P}$ and S$_{AP}$ are the fitted slopes as shown in Fig.2. $T_{fit}$ is the fitted total broadening at 0.2 K with low modulation voltage. }
 \begin{tabular}{ccccccc}
 \hline\hline
No. & M2 & M3 & M4  & M5 & A1 & A2     \\
\hline
R$_{P}$ (k$\Omega$) & 2.1 & 3 & 0.9 & 0.4 & 0.17 & 0.18   \\
R$_{AP}$ (k$\Omega$) & 10 & 12 & 3.6 &  1.1 & 0.26 & 0.28 \\
MR (\%) & 376 & 300 & 300 &  175 & 53 & 56  \\
S$_{P}$ & 0.0065 & 0.0064 & 0.0065 & 0.0037 & 0.003 & 0.0032  \\ 
S$_{AP}$ & 0.0113 & 0.0113 & 0.0113 & 0.0038 & 0.0039 & 0.004 \\ 
$T_{fit}^{P}$ (K) & & 0.7 & & 0.8 &  &   \\
$T_{fit}^{AP}$ (K) & & 0.8 & & 0.8 &  &  \\
\hline\hline
\end{tabular} 
\end{table}

First and second derivatives are recorded with an EG\&G 7265 digital lock-in amplifier with a home-made DC bias circuit.  Devices are cooled in a Leiden CF-450 cryogen free dilution refrigerator with a base temperature of 11 mK. Results of dV/dI and dI/dV measurements are checked for consistency. Note that the dip of $d^{2}V/d^{2}I$ measured here is proportional to a peak in $d^{2}I/d^{2}V$.~\cite{Holweg1992prb}

\section{Results and discussions}

\subsection{Differential resistance in a wide range}

Figure 1 shows the dV/dI for a typical device M3 at 3.6 K (similar results for AlO$_{x}$-based MTJs are shown in Appendix~\ref{appendix_AlO}).  Around 20-30 mV there is a broad `M' peak corresponding to interface magnon excitations; and around 80 mV a broad `Ph' peak corresponding to MgO phonon excitation.
In the P state the ZBA is very pronounced in the dV/dI plot since the background is almost flat, while in the AP state it is not as prominent. The background in the AP state increases almost linearly from -100 to -40 mV, and then forms a bulge between -40 to -15 mV. From IETS in Fig.~1b, it is clear that this bulge corresponds to the magnon emission, i.e., the `M' peak, although in the P state it is partially masked by the `ZB' peak at low bias. The `ZB' peak in the AP state in Fig.~1b is also easier to recognize than that in Fig.~1a. The above result suggests that to understand ZBA, we should focus on the bias range below the `M' peak and any broadening effect of the peak should be avoided. 

\begin{figure}
\includegraphics[width=9cm]{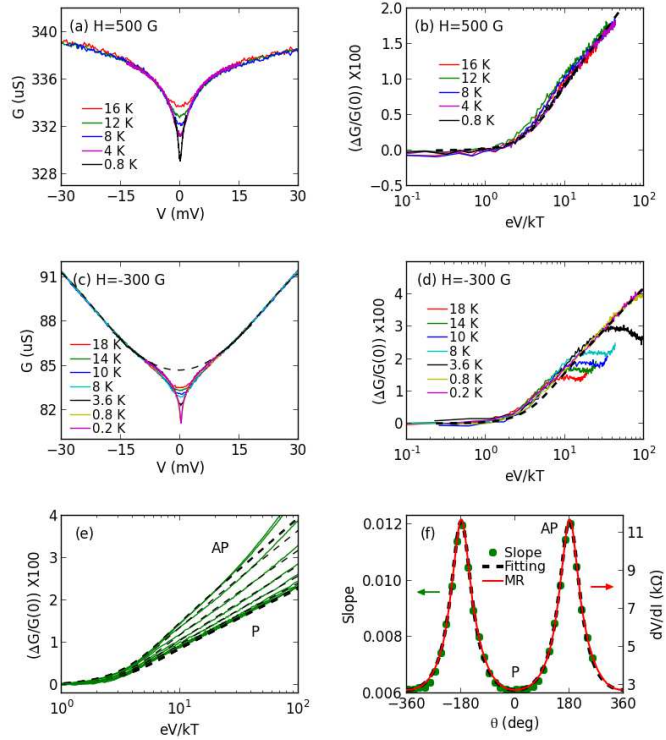}
\caption  {\small  (color online) Differential conductance vs bias for P state (a) and AP state (c), and $\Delta G$ vs $eV/k_{B}T$ on a log scale in the P (b) and AP (d) states, where $\Delta G(V)=G(V)-G(0)$. Curves in the P state are measured with 0.06 mV excitation voltage. An  effective temperature $T_{fit}$ = 1.2 K is assumed for the 0.8 K curve.  The curves in the AP state are measured with 0.1 mV excitation voltage. And $T_{fit}=$ 1.1 K and 0.7 K are assumed for the 0.8 K and 0.2 K data. $\Delta G$ curves with $\theta$ from 0$^{o}$ (P) to 180$^{o}$ (AP) at 0.2 K are shown in (e), and the angle dependence of fitted slopes and the resistance are shown in (f). Dashed lines in (b),(d), and (e) are fits using Eq.~(1).  Dashed line in (c) is a fit for background subtraction. Dashed line in (d) is a fit based on Eq.~(\ref{eqn_S_fit}). 
} 
\end{figure}

\subsection{Temperature dependence of ZBA}

The temperature dependence of the ZBA within $\pm$30 mV in both P and AP states are shown in Fig.~2a and 2c. We can see the conductance dip becomes deeper as temperature decreases. In Fig.~2b and 2d, the normalized conductance change  $\Delta G(V)/G(0)$, where $\Delta G(V)=G(V)-G(0)$, can be scaled on a single line with $\ln{(eV/k_{B}T)}$ when thermal broadening is small ($eV/k_{B}T> 1$). 
The scaling is good for temperatures from 16 K to 4 K. However, at 0.8 K, an elevated temperature $T_{fit}\sim$ 1.0 K needs to be assumed to overlap the data points with others at higher temperatures. And at even lower temperature,  $T_{fit}$ approaches a constant value (see also Table I). This discrepancy should be due to extra electromagnetic noises which has the same broadening effect as the modulation voltage. In this sense the ZBA serves as an internal electron thermometer similar to the case where EEI correction to conductance is used as an internal thermometer.~\cite{Wei2006} 

In the AP state, to remove the influence of the background especially at higher temperatures,  $\Delta G(V)/G(0)$   is derived after a background subtraction as denoted by a dashed line in Fig.~2c, which is a polynomial fit with points at $\pm 25$ mV, $\pm 20$ mV, and $\pm 15$ mV. The background may include the band structure effect~\cite{Cabrera2002apl} as well as the magnon contribution, and it is nearly flat at zero bias.~\cite{Simmons1963jap}  Then the normalized conductance change is 
$\frac{\Delta G(V)}{G(0)}=\frac{G'(V)-G'(0)}{G(0)}$,
where $G'(V)=G(V)-G_{B}(V)$, with the subscript `B' denotes background conductance. The scaling in Fig.~2d is clearly improved after background subtraction although for the curves at lowest temperatures it doesn't make much difference.  

Similar zero bias logarithmic singularity was previously observed for tunneling spectroscopy of various weakly disordered conductors.~\cite{Rowell1966prl,White1985prb} For aluminium films with different extent of disorder~\cite{Gershenzon1986jetp,Gershenzon1985jetpl} the ZBA can be well described by the EEI induced reduction of densisty of states $D$. The thermal broadening of the ZBA is similar to the broadening of spectral lines in tunneling spectroscopy of normal metals, which is about 5.4 $k_{B}T$ for inelastic tunneling.~\cite{Lambe1968pr} 

As explained in Ref.~[\onlinecite{Gershenzon1986jetp}] the thin film is considered as quasi 2D when $a\ll L_{T}=(\hbar \mathcal{D}/k_{B}T)^{1/2}$, where $L_{T}$ is the thermal diffusion length, and upper limit of the diffusion constant $\mathcal{D}$ for a thin film can be estimated as $v_{F}a/3$, and since $v_{F}\sim 10^{8}$ cm/sec, $a\sim 3$ nm, $\mathcal{D}\sim $ 10 cm$^{2}$/sec. So $L_{T} \sim 90(T/1 K)^{-1/2}$ nm, which is indeed much longer than $a$ below tens of K. In other words, the electrons interact with each other more frequently in reduced dimensionality when $a$ is smaller than the characteristic length scale defined by the EEI energy scale $k_{B}T$ as well as $eV$ (at higher bias there could be a transition from 2D to 3D). The change of tunneling conductance due to EEI is quantitatively described by the Altshuler-Aronov (AA) theory,~\cite{Altshuler1985,Gershenzon1986jetp}
\begin{equation}
\frac{G(V,T)-G(0,T)}{G(0,T)}=\frac{e^{2}R_{sq}}{8\pi^{2}\hbar}\ln{\frac{4\pi\delta}{\mathcal{D}R_{sq}}}[\Phi_{2}(\frac{eV}{k_{B}T})-\Phi_{2}(0)],
\label{Eq_AA}
\end{equation}
where  $R_{sq}$ is the resistance per square of the metal film, $\delta$ the thickness of the insulating barrier, $\mathcal{D}$ the diffusion constant ($\mathcal{D}$ is used to distinguish from the density of state $D$), and $\Phi_{2}$ a integral for 2D as defined in Ref.~[\onlinecite{Gershenzon1986jetp}]. The prefactor before the bracket can be lumped into one parameter $S$ and it is the only fitting parameter. When $eV\gg k_{B}T$, Eq.~(\ref{Eq_AA}) approaches $S\ln{\frac{eV}{k_{B}T}}$ and $S$ is just the slope shown in Fig.~2. Since $R_{sq}=\rho/a$, $a$ the thickness of the metal film, the resistivity $\rho=(e^{2}D \mathcal{D})^{-1}$, the slope $S \propto R_{sq}\ln(cD)$, where $c$ is a constant. To attempt a rough estimation of the slope we assume $\rho$ is about 400 $\mu\Omega\cdot cm$ in Ref.~[\onlinecite{Bernos2010prb}], and use $D$ of Fe which is estimated by electronic specific heat listed in textbook~\cite{Ashcroft1976} to be $1.8\times 10^{23} cm^{-3}eV^{-1}$. With these values the slope in Eq.~(\ref{Eq_AA}) is of the order of 0.01, close to the values listed in Table I.

The slopes observed in the P state (0.0065) and in the AP state (0.0113) are different for MTJs with MgO barrier, except for one device (M5) with low resistance and low MR (probably MgO barrier is thin and not fully crystallized). 
For device M5 both S$_{P}$ and S$_{AP}$ are close to 0.004, which can be put in the equation $\frac{1}{0.0065}+\frac{1}{0.0113}=\frac{1}{0.0041}$, as if $\Delta_{1}$ (dominating in P state) and $\Delta_{5}$ (dominating in AP state) channels are in parallel. The temperature dependence of resistance and the IETS all suggest M5 is a tunneling device and there is no pin hole. Thinner barrier may reduce the symmetry filtering effect as the bottom interface may not be perfect.~\cite{Teixeira2012apl}  Another observation is that for thin AlO$_{x}$-based MTJs, which is believed to be amorphous and has little symmetry filtering effect, S$_{AP}$ and S$_{P}$ are between 0.003 and 0.004, close to that of device M5. Since the AlO$_{x}$-based MTJs were not annealed, the property of CoFeB electrode is different than the case of MgO-based MTJs, so direct comparison is not feasible although the similarity seems not a coincidence.

In the crystalline MgO case the difference of $S_{P}$ and $S_{AP}$ can be formulated as follows (For detailed derivations see Appendix~\ref{appendix_model}). The conductance of tunneling junctions without considering thermal smearing (T=0) can be simplified as
\begin{eqnarray}
G_{\sigma\sigma',i}=\frac{e^2}{2h}\exp{(-\kappa_{i}\delta)}[D_{\sigma,i,L}(0)D_{\sigma',i,R}(-eV)+ \nonumber\\ 
D_{\sigma,i,L}(eV)D_{\sigma',i,R}(0)],\nonumber
\end{eqnarray}

where $\sigma$ is the majority ($\uparrow$) or minority  spin channels ($\downarrow$), $i$ the $\Delta_{1}$ ($i=1$) or $\Delta_{5}$ ($i=5$) symmetry channels (for simplicity we only consider $\Delta_{1}$ and $\Delta_{5}$ for $\uparrow$ and  $\Delta_{5}$ for $\downarrow$),  $L,R$ indicate the left and right electrodes. Considering symmetric junctions (left and right are indistinguishable), and assume the EEI theory is applicable for each symmetry channel independent of spin channel
\[
\frac{D_{i}(eV)-D_{i}(0)}{D_{i}(0)}=\frac{D_{i}(-eV)-D_{i}(0)}{D_{i}(0)}=s_{i}\ln{\frac{eV}{k_{B}T}}.
\]
Then the dependence of the slope on $\theta$, the angle between the magnetic moments in the two electrodes, $S(\theta)= $
\begin{equation}
s_{1}+(s_{5}-s_{1})\left( 
\frac{D_{1}(0)^{2}\exp{(-\kappa_{1}\delta)}}
 {2D_{5}(0)^{2}\exp{(-\kappa_{5}\delta)}} \cos(\frac{\theta}{2})^{2}+1\right)^{-1}
\label{eqn_S_fit}
\end{equation}
As shown in Fig.~2f, Eq.~(\ref{eqn_S_fit}) gives a good fit for both $S(\theta)$ and $R(\theta)$ with $\frac{D_{1}(0)^{2}\exp{(-\kappa_{1}\delta)}}{D_{5}(0)^{2}\exp{(-\kappa_{5}\delta)}}=13$, which suggests that the spin polarization of the tunneling current is 13/15=87\% in the P state.
For an amorphous barrier, the CoFe film is not crystallized in the (001) direction so that $k$ is not well defined. In a rough estimation we can assume randomization of the Bloch states, so $S_{P}\sim S_{AP}$.

\subsection{Comparison with previous investigations}

EEI was previously considered in Refs.~[\onlinecite{Nowak2000tsf}] and [\onlinecite{Bernos2010prb}] for  MgO and AlO-based MTJs, but in both cases a $\sqrt{V}$ dependence for 3D limit is used instead of 2D EEI used here (see Appendix~\ref{appendix_fit_previous} for comparison). 
Besides EEI, an alternative explanation for ZBA is magnetic impurity scattering inside barrier, presumably by diffused Mn or other impurities.~\cite{Drewello2009prb} No Mn was found in the barrier and this scenario is not consistent with the fact that after annealing ZBA is reduced.~\cite{Bernos2010prb,Wang2007prb} During the annealing process, the CoFeB layer partially converts to crystalline (001) CoFe with boron diffused away and this process reduces disorder ($R_{sq}$ is smaller), which leads to smaller ZBA following Eq.~(1). In the cases of epitaxial Fe/MgO/Fe MTJs~\cite{Ando2005apl,Du2010prb} and Co/MgO/Co MTJs,~\cite{Nishioka2008apl} the `ZB' peak is less pronounced suggesting that slight lattice mismatch does not introduce as much disorder compared with amorphous CoFeB. In addition, we found that magnetic impurity scattering does not fit data as well as EEI (cf. Appendix~\ref{appendix_fit_previous}).

\begin{figure}
\includegraphics[width=9cm]{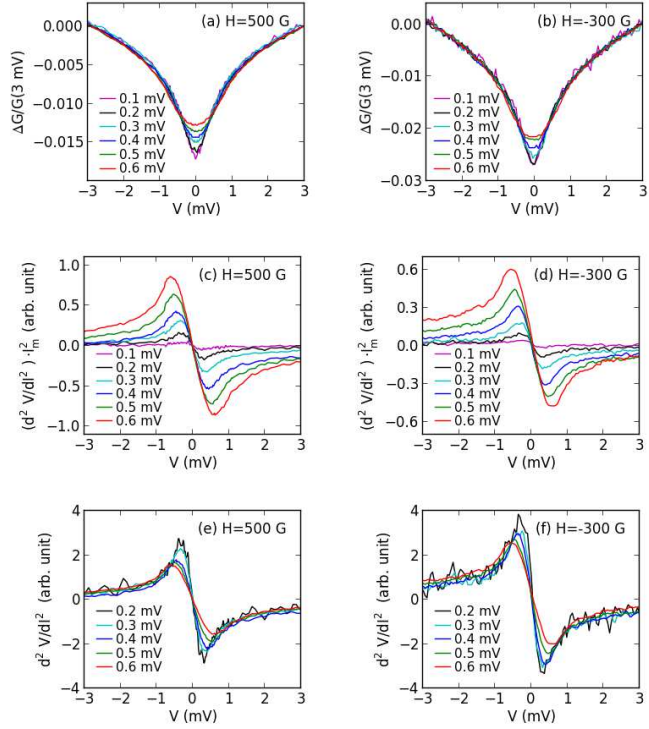}
\caption  {\small (color online)  Conductance change at 0.2 K in the P (a) and AP (b) states within $\pm$3 mV,  to emphasize the small change we redefine $\Delta G(V)=G(V)-G(3 mV)$. For IETS, second harmonic reading directly from lock-in ($(d^{2}V/dI^{2})\cdot I_{AC}^{2}$) are presented in the P (c) and AP (d) states, and $d^{2}V/dI^{2}$ (after dividing $I_{AC}^{2}$) in the P (e) and AP (f) states. Different modulation voltages are denoted with colors.  } 
\end{figure}

In previous investigations of MTJs the logarithmic dependence near zero bias can not be resolved because of large AC modulation voltage ($V_{m}$ about 2-4 mV in most cases) and high temperature ($\leq$ 10 K) smeared the features.~\cite{Drewello2009prb,Bernos2010prb}  The broadening effect due to AC modulation voltage is specifically shown in Fig.~3 at 0.2 K where thermal broadening is minimized.  Within a bias range of $\pm$3 mV, the conductance dip becomes deeper with lower $V_{m}$, from 0.6 mV to 0.1 mV, order of magnitude smaller than used in previous cases. The second derivatives in  Fig.~3e, 3f show that there is no fixed `ZB' point as presented in previous works, instead the `ZB' point can be infinitely close to zero with reduced modulation voltage if there is no other broadening effect. 
In fact, for IETS it is already known that the spectral broadening can be due to AC modulation voltage  $V_{m}$,~\cite{Klein1973prb,Wei2010prb} and the  experimental linewidth can be expressed as  $W_{exp}=(W_{intrinsic}^2+W_{thermal}^2+W_{m}^2)^{1/2}$,~\cite{Lauhon2001rsi,Reed2008mt} with $W_{thermal}=5.4k_{B}T/e$ and $W_{m}=1.22V_{m}$. Here below 0.8 K, the extra broadening indicated by the saturating $T_{fit}$ as the nominal temperature goes down should be due to extrinsic electromagnetic noises (line frequency noise etc). For AlO$_{x}$ MTJs with lower resistance, the pick up noise power is lower and $T_{fit}$ saturates at even lower temperature around 0.4 K. As mentioned earlier, this $T_{fit}$ can be considered as the real electronic temperature as calibrated by the quantum correction to density of states, similar to that calibrated by quantum correction to conductivity,~\cite{Wei2006} and that by the coulomb blockade effect.~\cite{Pekola1994prl}

\section{Conclusion}

In summary, in transport measurements of MTJs at low temperature and with low modulation voltage, logarithmic singularity near zero bias  is clearly observed and can be well described by the EEI theory in the weakly disordered quasi 2D limit. The commonly observed `ZB' peak in IETS is shown due to a broadening effect of this logarithmic singularity. While the slope of $\ln{(eV/k_{B}T)}$ dependence is almost the same for MTJs without symmetry filtering (amorphous barriers), different slopes are needed to fit the  $\ln{(eV/k_{B}T)}$ dependence in the P and AP states in high MR ratio MTJs, indicating that EEI of Bloch states with different symmetries can be distinguished. The angle dependence of the TMR and of the slope are also consistent. This finding may be useful to verify symmetry selective filtering for MTJs based on new Heusler alloys, and may also open a route for probing the EEI in correlated systems with electronic bands of different symmetries, e.g., SrRuO$_{3}$ where ``the degree and importance of correlation are still issues'',~\cite{Koster2012rmp}and La$_{1-x}$Sr$_{x}$MnO$_{3}$ as well as other perovskite oxides where metal-insulator transition and interface effects are still not fully understood.~\cite{Bhattacharya2008prl,Lepetit2012prl,Mazur2007prb}  Thus, similar ZBA experiments on SrRuO$_{3}$ and La$_{1-x}$Sr$_{x}$MnO$_{3}$ based MTJs~\cite{Herranz2003jap,Sun1998apl,ODonnell2000apl,Takahashi2003prb} can also be conducted.

\section*{Acknowledgements}

We thank Fa Wang for helpful discussions, particularly about correlated systems. Work supported by National Basic Research Program of China (973 Program) through Grant No. 2011CBA00106 and No. 2012CB927400, the State Key Project of Fundamental Research of Ministry of Science and Technology [MOST, No. 2010CB934401] and National Natural Science Foundation of China [NSFC, Grant No. 51229101]. A portion of this research was conducted at the Center for Nanophase Materials Sciences, sponsored at Oak Ridge National Laboratory by the Scientific User Facilities Division, Office of Basic Energy Sciences, U.S. Department of Energy.

\appendix

\section{Detailed results of AlO-based MTJ}
\label{appendix_AlO}

\begin{figure}[h]
\includegraphics[width=9cm]{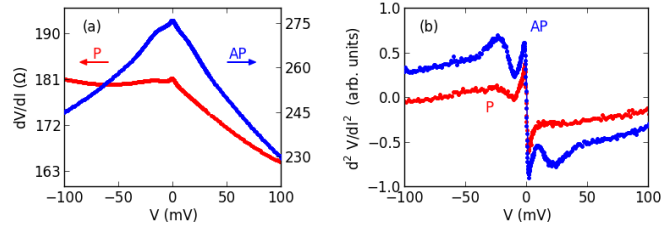}
\caption  {\small  (color online)  First and second derivatives at 3.6 K for sample A1. (a) Differential resistance in the P (red line) and AP  (blue line) states. (b) IETS in the P (red line, left coordinate) and AP (blue line, right coordinate) states. The maximum modulation voltage is about 0.5 mV.
\label{AlO_MTJ_IETS}
 }
\end{figure}

As listed in Table 1 of the main text, two AlO$_{x}$-based MTJ samples are measured. Figure~\ref{AlO_MTJ_IETS} shows the first and second derivatives for AlO$_{x}$-based MTJ sample A2 at 3.6 K. 
Asymmetric background is observed for the differential conductance and is more pronounced in the P state. Moreover, in Fig.~\ref{AlO_MTJ_IETS}b the `M' peak in the positive bias regime is much smaller than that in the negative bias regime. And the `ZB' peak is clearly visible in the IETS.

\begin{figure}
\includegraphics[width=9cm]{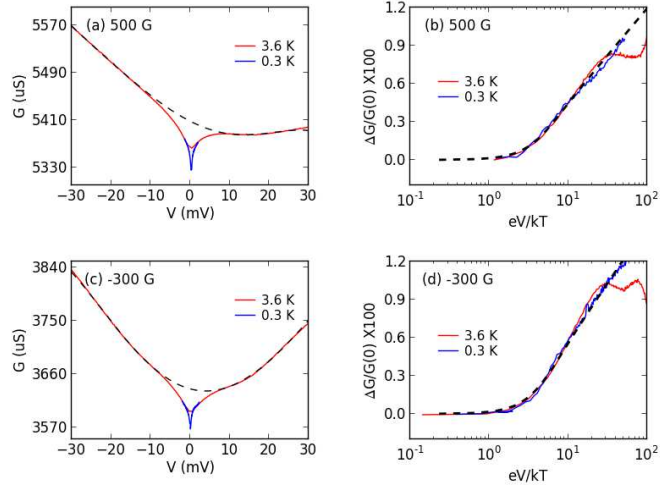}
\caption  {\small  (color online)  
Differential conductance vs bias voltage in the P (a) and AP (c) states, and normalized $\Delta G$ vs $eV/kT$  on a log scale in the P (b) and AP (d) states. The dashed lines in (a) and (c) are fittings for background subtraction. The dashed lines in (b) and (d) are fittings with 2D EEI theory.
\label{AlO_MTJ_fitting}
 }
\end{figure}

For AlO$_{x}$-based MTJs, $\Delta G(V)/G(0)$ is derived after a background subtraction in both the P and AP states, which are denoted by dashed lines with the $G(V)$ curves in Fig.~\ref{AlO_MTJ_fitting}a and Fig.~\ref{AlO_MTJ_fitting}c. The scaling with $\ln{(eV/kT)}$ and the EEI fittings are shown in Fig.~\ref{AlO_MTJ_fitting}b and Fig.~\ref{AlO_MTJ_fitting}d. It should be noted here that the fitted temperature $T_{fit}$=0.38 K for the nominal 0.3 K temperature, lower than $T_{fit}$=0.8 K for MgO-based MTJ measured at 0.2 K, which could be due to reduced extrinsic electromagnetic noise (line frequency etc) by better shielding, and due to lower resistance of the sample that also helps to reduce the noise voltage.

\section{A simple model for the fitted slopes}
\label{appendix_model}

Detailed modelling of tunneling can be found in monographs, e.g., see Ref.~[\onlinecite{Wolf2012}]. Here we use a simplified model, which neglects the off-normal incident electrons so that there is no integration over $k$, and write the current as an integration of the transmission coefficient over energy,
\begin{equation}
I=\frac{e}{h}\int_{\mu_{L}}^{\mu_{R}} \sum_{\sigma\sigma^{\prime}}T_{\sigma\sigma^{\prime}}(E) dE,
\end{equation}
where $\sigma$ is the majority ($\uparrow$) or minority ($\downarrow$) spin channels, L, R indicates the left and right electrodes, the bias energy $eV=\mu_L-\mu_R$. For simplicity, let us consider only two symmetry states, $\Delta_1$ and $\Delta_5$, and assume that the majority spin channel has both the $\Delta_1$ and $\Delta_5$ states, and the minority only the $\Delta_5$ state. The transmission for each symmetry channel is, 
\begin{equation}
T_{\sigma\sigma^{\prime},i}(E)=D_{\sigma,i,L}(E-\mu_{L})D_{\sigma^{\prime},i,R}(E-\mu_{R})\exp(-\kappa_i\delta),
\end{equation}
where $i$ is the $\Delta_1 (i=1)$ or $\Delta_1 (i=5)$ symmetry channels and $\delta$ is the barrier thickness. If one assumes that the ZBA arises from the $D$ factors, then the exponential factor due to the barrier can be treated as independent of the electron energy, so we have
\begin{equation}
I=\frac{e}{h}\sum_{\sigma\sigma^{\prime},i}\exp(-\kappa_{i}\delta)\int_{\mu_{L}}^{\mu_{R}} D_{\sigma,i,L}(E-\mu_{L})D_{\sigma^{\prime},i,R}(E-\mu_{R})dE.
\end{equation}
The conductance for each spin channel is
\begin{equation}
G_{\sigma\sigma^{\prime},i}=\frac{e^{2}}{2h}\exp(-\kappa_i\delta)[D_{\sigma,i,L}(0)D_{\sigma^{\prime},i,R}(-eV)
+D_{\sigma,i,L}(eV)D_{\sigma^{\prime},i,R}(0)].
\end{equation}
Considering symmetric junctions, and
assuming that the EEI theory is applicable for each symmetry
channel independent of spin,
\begin{equation}
\frac{D_i(eV)-D_i(0)}{D_i(0)}=\frac{D_i(-eV)-D_i(0)}{D_i(0)}=s_i\ln\frac{eV}{kT}.
\end{equation}
Then for P state, 
\begin{equation}
G_P(V)=G_{\uparrow\uparrow,1}(V)+G_{\uparrow\uparrow,5}(V)+G_{\downarrow\downarrow,5}(V),
\end{equation}
so that
$\frac{\Delta G_{P}}{G_{P}(0)}= $,
\begin{equation}
\frac{D_1(0)^2s_1\exp(-\kappa_1 \delta)+2D_5(0)^2s_5\exp(-\kappa_5 \delta)}{D_5(0)^2\exp(-\kappa_1 \delta)+2D_5(0)^2\exp(-\kappa_5 \delta)}\ln\frac{eV}{kT}
\approx s_1\ln(\frac{eV}{kT})
\end{equation}
where we assumed $D_{\uparrow5}=D_{\downarrow5}=D_5$. Similarly, for AP state only the $\Delta_5$ state can transmit,
\begin{equation}
G_{AP}(V)=2G_{\uparrow\downarrow,5}(V),
\end{equation}
so that
\begin{equation}
\frac{\Delta G_{AP}}{G_{AP}(0)}=s_5\ln(\frac{eV}{kT}).
\end{equation}
Clearly, if there is perfect spin filtering, the slopes of the P and AP conductance changes should be different.
 
If the magnetic moment in the two electrodes have an angle $\theta$, then to calculate the conductance we note that the spin part of the wave function from the two electrodes can be written as,
\begin{equation}
\Psi_{\uparrow L}=\Psi_{\uparrow R}\cos\frac{\theta}{2}+\Psi_{\downarrow R}\sin\frac{\theta}{2},
\end{equation}
and
\begin{equation}
\Psi_{\downarrow L}=-\Psi_{\uparrow R}\sin\frac{\theta}{2}+\Psi_{\downarrow R}\cos\frac{\theta}{2}.
\end{equation}
Thus,
\begin{eqnarray}
G(V)&=&\frac{e^2}{2h}\exp(-\kappa_1\delta)[D_1(0)D_1(-eV)
+D_1(eV)D_1(0)]\cos^2\frac{\theta}{2}\nonumber\\
&&+\frac{e^2}{h}\exp(-\kappa_5d)[D_5(0)D_5(-eV)+D_5(eV)D_5(0)],
\end{eqnarray}
The resistance is,
\begin{equation}
R=\frac{h}{e^2}\frac{1}{\exp(-\kappa_1\delta)[D_1(0)]^2\cos^2\frac{\theta}{2}+2\exp(-\kappa_5\delta)[D_5(0)]^2}.
\end{equation}
Then the $\theta$ dependence of the slope,
\begin{eqnarray}
S(\theta)&=&\frac{D_1(0)^2s_1\exp(-\kappa_1 \delta)\cos^2\frac{\theta}{2}+2D_5(0)^2s_5\exp(-\kappa_5 \delta)}{D_5(0)^2\exp(-\kappa_1 \delta)\cos^2\frac{\theta}{2}+2D_5(0)^2\exp(-\kappa_5 \delta)}\nonumber\\
&=&s_1+(s_5-s_1)[\frac{\exp(-\kappa_1\sigma)D_1(0)^2}{2\exp(-\kappa_5\sigma)D_5(0)^2}\cos^2\frac{\theta}{2}+1]^{-1}\nonumber\\
&=&s_1+\frac{2e^2}{h}\exp(-\kappa_5\delta)[D_5(0)]^2(s_5-s_1)R.
\end{eqnarray}
Therefore the slope varies linearly with the resistance as the angle is changed. This is confirmed experimentally, as shown in maintext with fitting parameters: $s_1=0.0052$,$s_5=0.0121$ and $\frac{\exp(-\kappa_1\delta)D_1(0)^2}{\exp(-\kappa_5\delta)D_5(0)^2}$=13, which also confirms the slope in the P state $S_P=\frac{13s_1+2s_5}{15}\approx s_1$.

The above simplified discussion can be made more realistic by including the $\Delta_2$ and $\Delta_{2\prime}$ states. The result will be more complicated but should not be qualitatively different.

\section{Fitting with magnetic impurity scattering}
\label{appendix_magnetic_impurity}

\begin{figure}
\includegraphics[width=9cm]{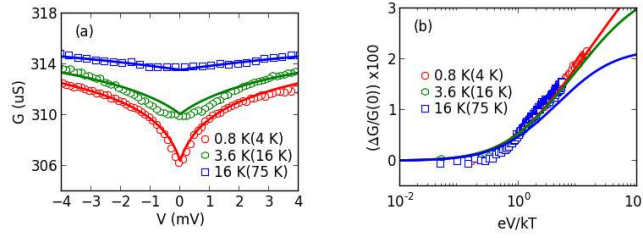}
\caption  {\small  (color online) (a) Conductance vs bias for sample M3. Solid lines are fits with Eq.~(\ref{eq:G_impur}). The fitted parameters are $E_{0}$=100 mV, $G_{2}$=321.5 uS, $G_{3}/G_{2}$=0.0076. (b) Normalized $\Delta G$ vs. $eV/kT$ for different temperatures. The temperatures in the parentheses are $T_{fit}$ assumed in Eq.~(\ref{eq:G_impur})
 } 
\label{fig:Mag_impur}
\end{figure}

Although magnetic impurity scattering is conventionally used to fit zero bias conductance peak,~\cite{Appelbaum1966prl} it has also been used to fit the peaks in the IETS as shown in Ref.~[\onlinecite{Wei2010prb}]. According to this model the conductance due to magnetic impurity scattering is
\begin{equation}
G_{impurity}=G_{2}-G_{3}F(|eV|,T),
\label{eq:G_impur}
\end{equation}
where $G_{2}$ can be viewed as a background conductance, and $G_{3}F$ yields the zero bias anomaly. The function $F(|eV|,T)$ is defined as
\begin{equation}
F(E,T)=\int_{-\infty}^{\infty} g(E^\prime,T)\frac{\partial}{\partial E^\prime}f(E^\prime-E) dE^\prime,
\end{equation}
with $f(E)$ the Fermi distribution function and the function
\begin{equation}
g(E,T)=\int_{-E_0}^{E_0}\frac{f(E^\prime)dE^\prime}{E^\prime-E}.
\end{equation}
An analytical approximation for $F(E,T)$ is\cite{Wei2010prb}
\begin{equation}
F(E,T)\approx\frac{\ln(1+\frac{E_0}{kT+E})}{1-\frac{kT}{E_0+0.4E}+\frac{12(kT)^2}{(E_0+2.4E)^2}}.
\label{eq:F}
\end{equation}
With this equation we can fit the data for MgO-based sample M3. In Fig.~\ref{fig:Mag_impur}a, elevated fitting temperatures 4 K, 16 K, 75 K are needed to fit the data at 0.8 K, 3.6 K and 16 K respectively. In Fig.~\ref{fig:Mag_impur}b, when $eV/kT>1$ Eq.~(\ref{eq:G_impur}) gives a linear dependence which bends down at larger bias, and the slopes are different for different temperatures, inconsistent with the experimental result. Here $E_{0}$=100 mV is used according to Ref.~[\onlinecite{Wei2010prb}] and $G_{3}$, $G_{2}$, and $T_{fit}$ are the fitting parameters. Moreover, in the range $1>E/kT>0.1$ the fitting curves are always slightly higher than the experimental data. In contrast, the 2D EEI model gives a straight line when $eV/kT>1$ that fits the data better and does not need higher $T_{fit}$ in this temperature range, and in the range $1>E/kT>0.1$ there is no deviation between the fit and the data as shown in Fig.2 in the main text.

To understand the discrepancy mentioned above, we note that $E_{0}$ determines the position where the curve bends down. Because there is no such bending down in experimental data, we may assume $E_{0}\gg E, kT$, so that
\begin{equation}
F(E,T)\approx-\ln(E+kT)+\ln(E_0),
\label{eq:F_approx}
\end{equation}
then
\begin{equation}
\frac{\Delta G}{G(0)}=\frac{1}{\frac{G_2}{G_3}-\ln(\frac{E_0}{kT})}\ln(\frac{E}{kT}+1),
\label{eq:G_approx}
\end{equation}
which is proportional to $\ln(\frac{E}{kT})$ when $E/kT\gg1$, similar to the of 2D EEI. But $\ln(\frac{E}{kT}+1)$ clearly deviates from $\ln(\frac{E}{kT})$ when $E/kT\leq1$. And for fixed $G_{3}/G_{2}$ the slope changes for different fitting temperatures as shown in Fig.~\ref{fig:Mag_impur}b.

Based on the discussion above, the 2D EEI model fits the data better than the magnetic impurity scattering model.


\section{2D EEI fit to data in previous work}
\label{appendix_fit_previous}

\begin{figure}
\includegraphics[width=9cm]{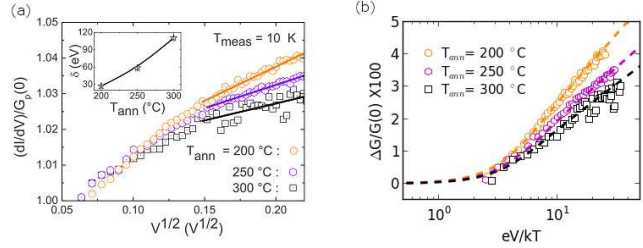}
\caption  {\small  (color online) (a) Normalized $G$ vs. $\sqrt{V}$ measured at 10 K for different T$_{ann}$ in the P state, copied from Ref.~[\onlinecite{Bernos2010prb}].  (b) The same data fitted with 2D EEI model with different T$_{fit}$.
 } 
 \label{fig:bernos}
\end{figure}

EEI was previously considered in Ref.~[\onlinecite{Bernos2010prb}]. In their study, the amplitude of the ZBA decreases with the annealing temperature T$_{ann}$ used to crystallize the CoFeB layers. And the fitting of ZBA in P state was done for the bias range between 23 and 50 mV with a $\sqrt{V}$ law applicable for 3D EEI. The original figure from Ref.~[\onlinecite{Bernos2010prb}] is reproduced in Fig.~\ref{fig:bernos}a. Then in Fig.~\ref{fig:bernos}b we show fitting using 2D EEI with the same data digitized from Fig.~\ref{fig:bernos}a, with fitting parameters listed in Table~\ref{table_I}. It is clear that 2D EEI gives better fits over a wider bias range.

The slopes in Table~\ref{table_I} are much larger than those for our MgO junctions (about 0.0065 for the P state), which could be explained by higher disorder at the interface of their junctions due to lower T$_{ann}$. For comparison,  T$_{ann}$ for our MTJs is 350 $^\circ$C and larger TMR are observed. In Table~\ref{table_I} different $T_{fit}$ higher than the 10 K nominal temperature are required for samples with different T$_{ann}$, which could be due to larger modulation voltages used in previous measurements.

\begin{table}
\caption{Fitting parameters  according to Eq.(1) in the main text for different annealing temperatures. $S_P$ is the fitted slope as shown in Fig.~\ref{fig:bernos}b. $T_{fit}$ is the fitted total broadening}
 \begin{tabular}{cccccccc}
 \hline\hline
T$_{ann}$($^\circ$C) \par & 200 & 250 & 300\\
\hline
S$_{P}$ & 0.018 & 0.014 & 0.012\\
T$_{fit}$(K) & 23 & 19 & 17\\
\hline\hline
\end{tabular} 
\label{table_I}
\end{table}


\begin{thebibliography}{67}%
\makeatletter
\providecommand \@ifxundefined [1]{%
 \@ifx{#1\undefined}
}%
\providecommand \@ifnum [1]{%
 \ifnum #1\expandafter \@firstoftwo
 \else \expandafter \@secondoftwo
 \fi
}%
\providecommand \@ifx [1]{%
 \ifx #1\expandafter \@firstoftwo
 \else \expandafter \@secondoftwo
 \fi
}%
\providecommand \natexlab [1]{#1}%
\providecommand \enquote  [1]{``#1''}%
\providecommand \bibnamefont  [1]{#1}%
\providecommand \bibfnamefont [1]{#1}%
\providecommand \citenamefont [1]{#1}%
\providecommand \href@noop [0]{\@secondoftwo}%
\providecommand \href [0]{\begingroup \@sanitize@url \@href}%
\providecommand \@href[1]{\@@startlink{#1}\@@href}%
\providecommand \@@href[1]{\endgroup#1\@@endlink}%
\providecommand \@sanitize@url [0]{\catcode `\\12\catcode `\$12\catcode
  `\&12\catcode `\#12\catcode `\^12\catcode `\_12\catcode `\%12\relax}%
\providecommand \@@startlink[1]{}%
\providecommand \@@endlink[0]{}%
\providecommand \url  [0]{\begingroup\@sanitize@url \@url }%
\providecommand \@url [1]{\endgroup\@href {#1}{\urlprefix }}%
\providecommand \urlprefix  [0]{URL }%
\providecommand \Eprint [0]{\href }%
\providecommand \doibase [0]{http://dx.doi.org/}%
\providecommand \selectlanguage [0]{\@gobble}%
\providecommand \bibinfo  [0]{\@secondoftwo}%
\providecommand \bibfield  [0]{\@secondoftwo}%
\providecommand \translation [1]{[#1]}%
\providecommand \BibitemOpen [0]{}%
\providecommand \bibitemStop [0]{}%
\providecommand \bibitemNoStop [0]{.\EOS\space}%
\providecommand \EOS [0]{\spacefactor3000\relax}%
\providecommand \BibitemShut  [1]{\csname bibitem#1\endcsname}%
\let\auto@bib@innerbib\@empty
\bibitem [{\citenamefont {Butler}\ \emph {et~al.}(2001)\citenamefont {Butler},
  \citenamefont {Zhang}, \citenamefont {Schulthess},\ and\ \citenamefont
  {MacLaren}}]{Butler2001prb}%
  \BibitemOpen
  \bibfield  {author} {\bibinfo {author} {\bibfnamefont {W.~H.}\ \bibnamefont
  {Butler}}, \bibinfo {author} {\bibfnamefont {X.-G.}\ \bibnamefont {Zhang}},
  \bibinfo {author} {\bibfnamefont {T.~C.}\ \bibnamefont {Schulthess}}, \ and\
  \bibinfo {author} {\bibfnamefont {J.~M.}\ \bibnamefont {MacLaren}},\ }\href
  {\doibase 10.1103/PhysRevB.63.054416} {\bibfield  {journal} {\bibinfo
  {journal} {Phys. Rev. B}\ }\textbf {\bibinfo {volume} {63}},\ \bibinfo
  {pages} {054416} (\bibinfo {year} {2001})}\BibitemShut {NoStop}%
\bibitem [{\citenamefont {Mathon}\ and\ \citenamefont
  {Umerski}(2001)}]{Mathon2001prb}%
  \BibitemOpen
  \bibfield  {author} {\bibinfo {author} {\bibfnamefont {J.}~\bibnamefont
  {Mathon}}\ and\ \bibinfo {author} {\bibfnamefont {A.}~\bibnamefont
  {Umerski}},\ }\href {\doibase 10.1103/PhysRevB.63.220403} {\bibfield
  {journal} {\bibinfo  {journal} {Phys. Rev. B}\ }\textbf {\bibinfo {volume}
  {63}},\ \bibinfo {pages} {220403} (\bibinfo {year} {2001})}\BibitemShut
  {NoStop}%
\bibitem [{\citenamefont {{Parkin}}\ \emph {et~al.}(2004)\citenamefont
  {{Parkin}}, \citenamefont {{Kaiser}}, \citenamefont {{Panchula}},
  \citenamefont {{Rice}}, \citenamefont {{Hughes}}, \citenamefont {{Samant}},\
  and\ \citenamefont {{Yang}}}]{Parkin2004nmat}%
  \BibitemOpen
  \bibfield  {author} {\bibinfo {author} {\bibfnamefont {S.~S.~P.}\
  \bibnamefont {{Parkin}}}, \bibinfo {author} {\bibfnamefont {C.}~\bibnamefont
  {{Kaiser}}}, \bibinfo {author} {\bibfnamefont {A.}~\bibnamefont
  {{Panchula}}}, \bibinfo {author} {\bibfnamefont {P.~M.}\ \bibnamefont
  {{Rice}}}, \bibinfo {author} {\bibfnamefont {B.}~\bibnamefont {{Hughes}}},
  \bibinfo {author} {\bibfnamefont {M.}~\bibnamefont {{Samant}}}, \ and\
  \bibinfo {author} {\bibfnamefont {S.-H.}\ \bibnamefont {{Yang}}},\ }\href
  {\doibase 10.1038/nmat1256} {\bibfield  {journal} {\bibinfo  {journal}
  {Nature Materials}\ }\textbf {\bibinfo {volume} {3}},\ \bibinfo {pages} {862}
  (\bibinfo {year} {2004})}\BibitemShut {NoStop}%
\bibitem [{\citenamefont {{Yuasa}}\ \emph {et~al.}(2004)\citenamefont
  {{Yuasa}}, \citenamefont {{Nagahama}}, \citenamefont {{Fukushima}},
  \citenamefont {{Suzuki}},\ and\ \citenamefont {{Ando}}}]{Yuasa2004nmat}%
  \BibitemOpen
  \bibfield  {author} {\bibinfo {author} {\bibfnamefont {S.}~\bibnamefont
  {{Yuasa}}}, \bibinfo {author} {\bibfnamefont {T.}~\bibnamefont {{Nagahama}}},
  \bibinfo {author} {\bibfnamefont {A.}~\bibnamefont {{Fukushima}}}, \bibinfo
  {author} {\bibfnamefont {Y.}~\bibnamefont {{Suzuki}}}, \ and\ \bibinfo
  {author} {\bibfnamefont {K.}~\bibnamefont {{Ando}}},\ }\href {\doibase
  10.1038/nmat1257} {\bibfield  {journal} {\bibinfo  {journal} {Nature
  Materials}\ }\textbf {\bibinfo {volume} {3}},\ \bibinfo {pages} {868}
  (\bibinfo {year} {2004})}\BibitemShut {NoStop}%
\bibitem [{\citenamefont {Zhang}\ and\ \citenamefont
  {Butler}(2004)}]{Zhang2004prb}%
  \BibitemOpen
  \bibfield  {author} {\bibinfo {author} {\bibfnamefont {X.-G.}\ \bibnamefont
  {Zhang}}\ and\ \bibinfo {author} {\bibfnamefont {W.~H.}\ \bibnamefont
  {Butler}},\ }\href {\doibase 10.1103/PhysRevB.70.172407} {\bibfield
  {journal} {\bibinfo  {journal} {Phys. Rev. B}\ }\textbf {\bibinfo {volume}
  {70}},\ \bibinfo {pages} {172407} (\bibinfo {year} {2004})}\BibitemShut
  {NoStop}%
\bibitem [{\citenamefont {Yuasa}\ and\ \citenamefont
  {Djayaprawira}(2007)}]{Yuasa2007jpd}%
  \BibitemOpen
  \bibfield  {author} {\bibinfo {author} {\bibfnamefont {S.}~\bibnamefont
  {Yuasa}}\ and\ \bibinfo {author} {\bibfnamefont {D.~D.}\ \bibnamefont
  {Djayaprawira}},\ }\href {http://stacks.iop.org/0022-3727/40/i=21/a=R01}
  {\bibfield  {journal} {\bibinfo  {journal} {Journal of Physics D: Applied
  Physics}\ }\textbf {\bibinfo {volume} {40}},\ \bibinfo {pages} {R337}
  (\bibinfo {year} {2007})}\BibitemShut {NoStop}%
\bibitem [{\citenamefont {Julliere}(1975)}]{Julliere1975}%
  \BibitemOpen
  \bibfield  {author} {\bibinfo {author} {\bibfnamefont {M.}~\bibnamefont
  {Julliere}},\ }\href {\doibase 10.1016/0375-9601(75)90174-7} {\bibfield
  {journal} {\bibinfo  {journal} {Phys. Lett. A}\ }\textbf {\bibinfo {volume}
  {54}},\ \bibinfo {pages} {225 } (\bibinfo {year} {1975})}\BibitemShut
  {NoStop}%
\bibitem [{\citenamefont {MacLaren}\ \emph {et~al.}(1997)\citenamefont
  {MacLaren}, \citenamefont {Zhang},\ and\ \citenamefont
  {Butler}}]{MacLaren1997prb}%
  \BibitemOpen
  \bibfield  {author} {\bibinfo {author} {\bibfnamefont {J.~M.}\ \bibnamefont
  {MacLaren}}, \bibinfo {author} {\bibfnamefont {X.-G.}\ \bibnamefont {Zhang}},
  \ and\ \bibinfo {author} {\bibfnamefont {W.~H.}\ \bibnamefont {Butler}},\
  }\href {\doibase 10.1103/PhysRevB.56.11827} {\bibfield  {journal} {\bibinfo
  {journal} {Phys. Rev. B}\ }\textbf {\bibinfo {volume} {56}},\ \bibinfo
  {pages} {11827} (\bibinfo {year} {1997})}\BibitemShut {NoStop}%
\bibitem [{\citenamefont {Miao}\ \emph {et~al.}(2011)\citenamefont {Miao},
  \citenamefont {M\"{u}nzenberg},\ and\ \citenamefont {Moodera}}]{Miao2011rpp}%
  \BibitemOpen
  \bibfield  {author} {\bibinfo {author} {\bibfnamefont {G.-X.}\ \bibnamefont
  {Miao}}, \bibinfo {author} {\bibfnamefont {M.}~\bibnamefont
  {M\"{u}nzenberg}}, \ and\ \bibinfo {author} {\bibfnamefont {J.~S.}\
  \bibnamefont {Moodera}},\ }\href
  {http://stacks.iop.org/0034-4885/74/i=3/a=036501} {\bibfield  {journal}
  {\bibinfo  {journal} {Reports on Progress in Physics}\ }\textbf {\bibinfo
  {volume} {74}},\ \bibinfo {pages} {036501} (\bibinfo {year}
  {2011})}\BibitemShut {NoStop}%
\bibitem [{\citenamefont {BAI}\ \emph {et~al.}(2012)\citenamefont {BAI},
  \citenamefont {SHEN}, \citenamefont {HAN},\ and\ \citenamefont
  {FENG}}]{Bai2012spin}%
  \BibitemOpen
  \bibfield  {author} {\bibinfo {author} {\bibfnamefont {Z.}~\bibnamefont
  {BAI}}, \bibinfo {author} {\bibfnamefont {L.}~\bibnamefont {SHEN}}, \bibinfo
  {author} {\bibfnamefont {G.}~\bibnamefont {HAN}}, \ and\ \bibinfo {author}
  {\bibfnamefont {Y.~P.}\ \bibnamefont {FENG}},\ }\href {\doibase
  10.1142/S201032471230006X} {\bibfield  {journal} {\bibinfo  {journal} {SPIN}\
  }\textbf {\bibinfo {volume} {02}},\ \bibinfo {pages} {1230006} (\bibinfo
  {year} {2012})}\BibitemShut {NoStop}%
\bibitem [{\citenamefont {Sukegawa}\ \emph {et~al.}(2009)\citenamefont
  {Sukegawa}, \citenamefont {Wang}, \citenamefont {Shan}, \citenamefont
  {Nakatani}, \citenamefont {Inomata},\ and\ \citenamefont
  {Hono}}]{Sukegawa2009prb}%
  \BibitemOpen
  \bibfield  {author} {\bibinfo {author} {\bibfnamefont {H.}~\bibnamefont
  {Sukegawa}}, \bibinfo {author} {\bibfnamefont {W.}~\bibnamefont {Wang}},
  \bibinfo {author} {\bibfnamefont {R.}~\bibnamefont {Shan}}, \bibinfo {author}
  {\bibfnamefont {T.}~\bibnamefont {Nakatani}}, \bibinfo {author}
  {\bibfnamefont {K.}~\bibnamefont {Inomata}}, \ and\ \bibinfo {author}
  {\bibfnamefont {K.}~\bibnamefont {Hono}},\ }\href {\doibase
  10.1103/PhysRevB.79.184418} {\bibfield  {journal} {\bibinfo  {journal} {Phys.
  Rev. B}\ }\textbf {\bibinfo {volume} {79}},\ \bibinfo {pages} {184418}
  (\bibinfo {year} {2009})}\BibitemShut {NoStop}%
\bibitem [{\citenamefont {Wang}\ \emph {et~al.}(2009)\citenamefont {Wang},
  \citenamefont {Sukegawa}, \citenamefont {Shan}, \citenamefont {Mitani},\ and\
  \citenamefont {Inomata}}]{Wang2009apl}%
  \BibitemOpen
  \bibfield  {author} {\bibinfo {author} {\bibfnamefont {W.}~\bibnamefont
  {Wang}}, \bibinfo {author} {\bibfnamefont {H.}~\bibnamefont {Sukegawa}},
  \bibinfo {author} {\bibfnamefont {R.}~\bibnamefont {Shan}}, \bibinfo {author}
  {\bibfnamefont {S.}~\bibnamefont {Mitani}}, \ and\ \bibinfo {author}
  {\bibfnamefont {K.}~\bibnamefont {Inomata}},\ }\href {\doibase
  10.1063/1.3258069} {\bibfield  {journal} {\bibinfo  {journal} {Appl. Phys.
  Lett.}\ }\textbf {\bibinfo {volume} {95}},\ \bibinfo {eid} {182502} (\bibinfo
  {year} {2009})}\BibitemShut {NoStop}%
\bibitem [{\citenamefont {{Jourdan}}\ \emph {et~al.}(2014)\citenamefont
  {{Jourdan}}, \citenamefont {{Min{\'a}r}}, \citenamefont {{Braun}},
  \citenamefont {{Kronenberg}}, \citenamefont {{Chadov}}, \citenamefont
  {{Balke}}, \citenamefont {{Gloskovskii}}, \citenamefont {{Kolbe}},
  \citenamefont {{Elmers}}, \citenamefont {{Sch{\"o}nhense}}, \citenamefont
  {{Ebert}}, \citenamefont {{Felser}},\ and\ \citenamefont
  {{Kl{\"a}ui}}}]{Jourdan2014ncomms}%
  \BibitemOpen
  \bibfield  {author} {\bibinfo {author} {\bibfnamefont {M.}~\bibnamefont
  {{Jourdan}}}, \bibinfo {author} {\bibfnamefont {J.}~\bibnamefont
  {{Min{\'a}r}}}, \bibinfo {author} {\bibfnamefont {J.}~\bibnamefont
  {{Braun}}}, \bibinfo {author} {\bibfnamefont {A.}~\bibnamefont
  {{Kronenberg}}}, \bibinfo {author} {\bibfnamefont {S.}~\bibnamefont
  {{Chadov}}}, \bibinfo {author} {\bibfnamefont {B.}~\bibnamefont {{Balke}}},
  \bibinfo {author} {\bibfnamefont {A.}~\bibnamefont {{Gloskovskii}}}, \bibinfo
  {author} {\bibfnamefont {M.}~\bibnamefont {{Kolbe}}}, \bibinfo {author}
  {\bibfnamefont {H.~J.}\ \bibnamefont {{Elmers}}}, \bibinfo {author}
  {\bibfnamefont {G.}~\bibnamefont {{Sch{\"o}nhense}}}, \bibinfo {author}
  {\bibfnamefont {H.}~\bibnamefont {{Ebert}}}, \bibinfo {author} {\bibfnamefont
  {C.}~\bibnamefont {{Felser}}}, \ and\ \bibinfo {author} {\bibfnamefont
  {M.}~\bibnamefont {{Kl{\"a}ui}}},\ }\href {\doibase 10.1038/ncomms4974}
  {\bibfield  {journal} {\bibinfo  {journal} {Nature Communications}\ }\textbf
  {\bibinfo {volume} {5}},\ \bibinfo {eid} {3974} (\bibinfo {year} {2014}),\
  10.1038/ncomms4974}\BibitemShut {NoStop}%
\bibitem [{\citenamefont {Wang}\ \emph {et~al.}(2010)\citenamefont {Wang},
  \citenamefont {Liu}, \citenamefont {Kodzuka}, \citenamefont {Sukegawa},
  \citenamefont {Wojcik}, \citenamefont {Jedryka}, \citenamefont {Wu},
  \citenamefont {Inomata}, \citenamefont {Mitani},\ and\ \citenamefont
  {Hono}}]{Wang2010prb_Coherent}%
  \BibitemOpen
  \bibfield  {author} {\bibinfo {author} {\bibfnamefont {W.}~\bibnamefont
  {Wang}}, \bibinfo {author} {\bibfnamefont {E.}~\bibnamefont {Liu}}, \bibinfo
  {author} {\bibfnamefont {M.}~\bibnamefont {Kodzuka}}, \bibinfo {author}
  {\bibfnamefont {H.}~\bibnamefont {Sukegawa}}, \bibinfo {author}
  {\bibfnamefont {M.}~\bibnamefont {Wojcik}}, \bibinfo {author} {\bibfnamefont
  {E.}~\bibnamefont {Jedryka}}, \bibinfo {author} {\bibfnamefont {G.~H.}\
  \bibnamefont {Wu}}, \bibinfo {author} {\bibfnamefont {K.}~\bibnamefont
  {Inomata}}, \bibinfo {author} {\bibfnamefont {S.}~\bibnamefont {Mitani}}, \
  and\ \bibinfo {author} {\bibfnamefont {K.}~\bibnamefont {Hono}},\ }\href
  {\doibase 10.1103/PhysRevB.81.140402} {\bibfield  {journal} {\bibinfo
  {journal} {Phys. Rev. B}\ }\textbf {\bibinfo {volume} {81}},\ \bibinfo
  {pages} {140402} (\bibinfo {year} {2010})}\BibitemShut {NoStop}%
\bibitem [{\citenamefont {Zhang}\ \emph {et~al.}(2008)\citenamefont {Zhang},
  \citenamefont {Wang},\ and\ \citenamefont {Han}}]{Zhang2008prb}%
  \BibitemOpen
  \bibfield  {author} {\bibinfo {author} {\bibfnamefont {X.-G.}\ \bibnamefont
  {Zhang}}, \bibinfo {author} {\bibfnamefont {Y.}~\bibnamefont {Wang}}, \ and\
  \bibinfo {author} {\bibfnamefont {X.~F.}\ \bibnamefont {Han}},\ }\href
  {\doibase 10.1103/PhysRevB.77.144431} {\bibfield  {journal} {\bibinfo
  {journal} {Phys. Rev. B}\ }\textbf {\bibinfo {volume} {77}},\ \bibinfo
  {pages} {144431} (\bibinfo {year} {2008})}\BibitemShut {NoStop}%
\bibitem [{\citenamefont {Rowell}\ and\ \citenamefont
  {Shen}(1966)}]{Rowell1966prl}%
  \BibitemOpen
  \bibfield  {author} {\bibinfo {author} {\bibfnamefont {J.~M.}\ \bibnamefont
  {Rowell}}\ and\ \bibinfo {author} {\bibfnamefont {L.~Y.~L.}\ \bibnamefont
  {Shen}},\ }\href {\doibase 10.1103/PhysRevLett.17.15} {\bibfield  {journal}
  {\bibinfo  {journal} {Phys. Rev. Lett.}\ }\textbf {\bibinfo {volume} {17}},\
  \bibinfo {pages} {15} (\bibinfo {year} {1966})}\BibitemShut {NoStop}%
\bibitem [{\citenamefont {Appelbaum}(1967)}]{Appelbaum1967pr}%
  \BibitemOpen
  \bibfield  {author} {\bibinfo {author} {\bibfnamefont {J.~A.}\ \bibnamefont
  {Appelbaum}},\ }\href {\doibase 10.1103/PhysRev.154.633} {\bibfield
  {journal} {\bibinfo  {journal} {Phys. Rev.}\ }\textbf {\bibinfo {volume}
  {154}},\ \bibinfo {pages} {633} (\bibinfo {year} {1967})}\BibitemShut
  {NoStop}%
\bibitem [{\citenamefont {Appelbaum}(1966)}]{Appelbaum1966prl}%
  \BibitemOpen
  \bibfield  {author} {\bibinfo {author} {\bibfnamefont {J.}~\bibnamefont
  {Appelbaum}},\ }\href {\doibase 10.1103/PhysRevLett.17.91} {\bibfield
  {journal} {\bibinfo  {journal} {Phys. Rev. Lett.}\ }\textbf {\bibinfo
  {volume} {17}},\ \bibinfo {pages} {91} (\bibinfo {year} {1966})}\BibitemShut
  {NoStop}%
\bibitem [{\citenamefont {Shen}\ and\ \citenamefont
  {Rowell}(1968)}]{Shen1968pr}%
  \BibitemOpen
  \bibfield  {author} {\bibinfo {author} {\bibfnamefont {L.~Y.~L.}\
  \bibnamefont {Shen}}\ and\ \bibinfo {author} {\bibfnamefont {J.~M.}\
  \bibnamefont {Rowell}},\ }\href {\doibase 10.1103/PhysRev.165.566} {\bibfield
   {journal} {\bibinfo  {journal} {Phys. Rev.}\ }\textbf {\bibinfo {volume}
  {165}},\ \bibinfo {pages} {566} (\bibinfo {year} {1968})}\BibitemShut
  {NoStop}%
\bibitem [{\citenamefont {Gloos}(2009)}]{Gloos2009ltp}%
  \BibitemOpen
  \bibfield  {author} {\bibinfo {author} {\bibfnamefont {K.}~\bibnamefont
  {Gloos}},\ }\href {\doibase 10.1063/1.3274810} {\bibfield  {journal}
  {\bibinfo  {journal} {Low Temperature Physics}\ }\textbf {\bibinfo {volume}
  {35}},\ \bibinfo {pages} {935} (\bibinfo {year} {2009})}\BibitemShut
  {NoStop}%
\bibitem [{\citenamefont {Moodera}\ \emph {et~al.}(1995)\citenamefont
  {Moodera}, \citenamefont {Kinder}, \citenamefont {Wong},\ and\ \citenamefont
  {Meservey}}]{Moodera1995prl}%
  \BibitemOpen
  \bibfield  {author} {\bibinfo {author} {\bibfnamefont {J.~S.}\ \bibnamefont
  {Moodera}}, \bibinfo {author} {\bibfnamefont {L.~R.}\ \bibnamefont {Kinder}},
  \bibinfo {author} {\bibfnamefont {T.~M.}\ \bibnamefont {Wong}}, \ and\
  \bibinfo {author} {\bibfnamefont {R.}~\bibnamefont {Meservey}},\ }\href
  {\doibase 10.1103/PhysRevLett.74.3273} {\bibfield  {journal} {\bibinfo
  {journal} {Phys. Rev. Lett.}\ }\textbf {\bibinfo {volume} {74}},\ \bibinfo
  {pages} {3273} (\bibinfo {year} {1995})}\BibitemShut {NoStop}%
\bibitem [{\citenamefont {Gallagher}\ \emph {et~al.}(1997)\citenamefont
  {Gallagher}, \citenamefont {Parkin}, \citenamefont {Lu}, \citenamefont
  {Bian}, \citenamefont {Marley}, \citenamefont {Roche}, \citenamefont
  {Altman}, \citenamefont {Rishton}, \citenamefont {Jahnes}, \citenamefont
  {Shaw},\ and\ \citenamefont {Xiao}}]{Gallagher1997jap}%
  \BibitemOpen
  \bibfield  {author} {\bibinfo {author} {\bibfnamefont {W.~J.}\ \bibnamefont
  {Gallagher}}, \bibinfo {author} {\bibfnamefont {S.~S.~P.}\ \bibnamefont
  {Parkin}}, \bibinfo {author} {\bibfnamefont {Y.}~\bibnamefont {Lu}}, \bibinfo
  {author} {\bibfnamefont {X.~P.}\ \bibnamefont {Bian}}, \bibinfo {author}
  {\bibfnamefont {A.}~\bibnamefont {Marley}}, \bibinfo {author} {\bibfnamefont
  {K.~P.}\ \bibnamefont {Roche}}, \bibinfo {author} {\bibfnamefont {R.~A.}\
  \bibnamefont {Altman}}, \bibinfo {author} {\bibfnamefont {S.~A.}\
  \bibnamefont {Rishton}}, \bibinfo {author} {\bibfnamefont {C.}~\bibnamefont
  {Jahnes}}, \bibinfo {author} {\bibfnamefont {T.~M.}\ \bibnamefont {Shaw}}, \
  and\ \bibinfo {author} {\bibfnamefont {G.}~\bibnamefont {Xiao}},\ }\href
  {\doibase 10.1063/1.364744} {\bibfield  {journal} {\bibinfo  {journal} {J.
  Appl. Phys.}\ }\textbf {\bibinfo {volume} {81}},\ \bibinfo {pages} {3741}
  (\bibinfo {year} {1997})}\BibitemShut {NoStop}%
\bibitem [{\citenamefont {Matsumoto}\ \emph {et~al.}(2005)\citenamefont
  {Matsumoto}, \citenamefont {Hamada}, \citenamefont {Mizuguchi}, \citenamefont
  {Shiraishi}, \citenamefont {Maehara}, \citenamefont {Tsunekawa},
  \citenamefont {Djayaprawira}, \citenamefont {Watanabe}, \citenamefont
  {Kurosaki}, \citenamefont {Nagahama}, \citenamefont {Fukushima},
  \citenamefont {Kubota}, \citenamefont {Yuasa},\ and\ \citenamefont
  {Suzuki}}]{Matsumoto2005ssc}%
  \BibitemOpen
  \bibfield  {author} {\bibinfo {author} {\bibfnamefont {R.}~\bibnamefont
  {Matsumoto}}, \bibinfo {author} {\bibfnamefont {Y.}~\bibnamefont {Hamada}},
  \bibinfo {author} {\bibfnamefont {M.}~\bibnamefont {Mizuguchi}}, \bibinfo
  {author} {\bibfnamefont {M.}~\bibnamefont {Shiraishi}}, \bibinfo {author}
  {\bibfnamefont {H.}~\bibnamefont {Maehara}}, \bibinfo {author} {\bibfnamefont
  {K.}~\bibnamefont {Tsunekawa}}, \bibinfo {author} {\bibfnamefont
  {D.}~\bibnamefont {Djayaprawira}}, \bibinfo {author} {\bibfnamefont
  {N.}~\bibnamefont {Watanabe}}, \bibinfo {author} {\bibfnamefont
  {Y.}~\bibnamefont {Kurosaki}}, \bibinfo {author} {\bibfnamefont
  {T.}~\bibnamefont {Nagahama}}, \bibinfo {author} {\bibfnamefont
  {A.}~\bibnamefont {Fukushima}}, \bibinfo {author} {\bibfnamefont
  {H.}~\bibnamefont {Kubota}}, \bibinfo {author} {\bibfnamefont
  {S.}~\bibnamefont {Yuasa}}, \ and\ \bibinfo {author} {\bibfnamefont
  {Y.}~\bibnamefont {Suzuki}},\ }\href {\doibase 10.1016/j.ssc.2005.08.032}
  {\bibfield  {journal} {\bibinfo  {journal} {Solid State Commun.}\ }\textbf
  {\bibinfo {volume} {136}},\ \bibinfo {pages} {611} (\bibinfo {year}
  {2005})}\BibitemShut {NoStop}%
\bibitem [{\citenamefont {Ikegawa}\ \emph {et~al.}(2007)\citenamefont
  {Ikegawa}, \citenamefont {Aikawa}, \citenamefont {Ueda}, \citenamefont
  {Nagamine}, \citenamefont {Shimomura}, \citenamefont {Yoshikawa},
  \citenamefont {Hosotani},\ and\ \citenamefont {Yoda}}]{Ikegawa2007jap}%
  \BibitemOpen
  \bibfield  {author} {\bibinfo {author} {\bibfnamefont {S.}~\bibnamefont
  {Ikegawa}}, \bibinfo {author} {\bibfnamefont {H.}~\bibnamefont {Aikawa}},
  \bibinfo {author} {\bibfnamefont {T.}~\bibnamefont {Ueda}}, \bibinfo {author}
  {\bibfnamefont {M.}~\bibnamefont {Nagamine}}, \bibinfo {author}
  {\bibfnamefont {N.}~\bibnamefont {Shimomura}}, \bibinfo {author}
  {\bibfnamefont {M.}~\bibnamefont {Yoshikawa}}, \bibinfo {author}
  {\bibfnamefont {K.}~\bibnamefont {Hosotani}}, \ and\ \bibinfo {author}
  {\bibfnamefont {H.}~\bibnamefont {Yoda}},\ }\href@noop {} {\bibfield
  {journal} {\bibinfo  {journal} {J. Appl. Phys.}\ }\textbf {\bibinfo {volume}
  {101}},\ \bibinfo {eid} {09} (\bibinfo {year} {2007})}\BibitemShut {NoStop}%
\bibitem [{\citenamefont {Zhang}\ \emph {et~al.}(1997)\citenamefont {Zhang},
  \citenamefont {Levy}, \citenamefont {Marley},\ and\ \citenamefont
  {Parkin}}]{Zhang1997prl}%
  \BibitemOpen
  \bibfield  {author} {\bibinfo {author} {\bibfnamefont {S.}~\bibnamefont
  {Zhang}}, \bibinfo {author} {\bibfnamefont {P.~M.}\ \bibnamefont {Levy}},
  \bibinfo {author} {\bibfnamefont {A.~C.}\ \bibnamefont {Marley}}, \ and\
  \bibinfo {author} {\bibfnamefont {S.~S.~P.}\ \bibnamefont {Parkin}},\ }\href
  {\doibase 10.1103/PhysRevLett.79.3744} {\bibfield  {journal} {\bibinfo
  {journal} {Phys. Rev. Lett.}\ }\textbf {\bibinfo {volume} {79}},\ \bibinfo
  {pages} {3744} (\bibinfo {year} {1997})}\BibitemShut {NoStop}%
\bibitem [{\citenamefont {Cabrera}\ and\ \citenamefont
  {Garc{\'i}a}(2002)}]{Cabrera2002apl}%
  \BibitemOpen
  \bibfield  {author} {\bibinfo {author} {\bibfnamefont {G.~G.}\ \bibnamefont
  {Cabrera}}\ and\ \bibinfo {author} {\bibfnamefont {N.}~\bibnamefont
  {Garc{\'i}a}},\ }\href {\doibase 10.1063/1.1433168} {\bibfield  {journal}
  {\bibinfo  {journal} {Appl. Phys. Lett.}\ }\textbf {\bibinfo {volume} {80}},\
  \bibinfo {pages} {1782} (\bibinfo {year} {2002})}\BibitemShut {NoStop}%
\bibitem [{\citenamefont {Bernos}\ \emph {et~al.}(2010)\citenamefont {Bernos},
  \citenamefont {Hehn}, \citenamefont {Montaigne}, \citenamefont {Tiusan},
  \citenamefont {Lacour}, \citenamefont {Alnot}, \citenamefont {Negulescu},
  \citenamefont {Lengaigne}, \citenamefont {Snoeck},\ and\ \citenamefont
  {Aliev}}]{Bernos2010prb}%
  \BibitemOpen
  \bibfield  {author} {\bibinfo {author} {\bibfnamefont {J.}~\bibnamefont
  {Bernos}}, \bibinfo {author} {\bibfnamefont {M.}~\bibnamefont {Hehn}},
  \bibinfo {author} {\bibfnamefont {F.}~\bibnamefont {Montaigne}}, \bibinfo
  {author} {\bibfnamefont {C.}~\bibnamefont {Tiusan}}, \bibinfo {author}
  {\bibfnamefont {D.}~\bibnamefont {Lacour}}, \bibinfo {author} {\bibfnamefont
  {M.}~\bibnamefont {Alnot}}, \bibinfo {author} {\bibfnamefont
  {B.}~\bibnamefont {Negulescu}}, \bibinfo {author} {\bibfnamefont
  {G.}~\bibnamefont {Lengaigne}}, \bibinfo {author} {\bibfnamefont
  {E.}~\bibnamefont {Snoeck}}, \ and\ \bibinfo {author} {\bibfnamefont {F.~G.}\
  \bibnamefont {Aliev}},\ }\href {\doibase 10.1103/PhysRevB.82.060405}
  {\bibfield  {journal} {\bibinfo  {journal} {Phys. Rev. B}\ }\textbf {\bibinfo
  {volume} {82}},\ \bibinfo {pages} {060405} (\bibinfo {year}
  {2010})}\BibitemShut {NoStop}%
\bibitem [{\citenamefont {{Nowak}}(2000)}]{Nowak2000tsf}%
  \BibitemOpen
  \bibfield  {author} {\bibinfo {author} {\bibfnamefont {E.}~\bibnamefont
  {{Nowak}}},\ }\href {\doibase 10.1016/S0040-6090(00)01284-0} {\bibfield
  {journal} {\bibinfo  {journal} {Thin Solid Films}\ }\textbf {\bibinfo
  {volume} {377}},\ \bibinfo {pages} {699} (\bibinfo {year}
  {2000})}\BibitemShut {NoStop}%
\bibitem [{\citenamefont {Han}\ \emph {et~al.}(2001)\citenamefont {Han},
  \citenamefont {Murai}, \citenamefont {Ando}, \citenamefont {Kubota},\ and\
  \citenamefont {Miyazaki}}]{Han2001apl}%
  \BibitemOpen
  \bibfield  {author} {\bibinfo {author} {\bibfnamefont {X.-F.}\ \bibnamefont
  {Han}}, \bibinfo {author} {\bibfnamefont {J.}~\bibnamefont {Murai}}, \bibinfo
  {author} {\bibfnamefont {Y.}~\bibnamefont {Ando}}, \bibinfo {author}
  {\bibfnamefont {H.}~\bibnamefont {Kubota}}, \ and\ \bibinfo {author}
  {\bibfnamefont {T.}~\bibnamefont {Miyazaki}},\ }\href {\doibase
  10.1063/1.1367882} {\bibfield  {journal} {\bibinfo  {journal} {Appl. Phys.
  Lett.}\ }\textbf {\bibinfo {volume} {78}},\ \bibinfo {pages} {2533} (\bibinfo
  {year} {2001})}\BibitemShut {NoStop}%
\bibitem [{\citenamefont {Bang}\ \emph {et~al.}(2009)\citenamefont {Bang},
  \citenamefont {Nozaki}, \citenamefont {Djayaprawira}, \citenamefont
  {Shiraishi}, \citenamefont {Suzuki}, \citenamefont {Fukushima}, \citenamefont
  {Kubota}, \citenamefont {Nagahama}, \citenamefont {Yuasa}, \citenamefont
  {Maehara}, \citenamefont {Tsunekawa}, \citenamefont {Nagamine}, \citenamefont
  {Watanabe},\ and\ \citenamefont {Itoh}}]{Bang2009jap}%
  \BibitemOpen
  \bibfield  {author} {\bibinfo {author} {\bibfnamefont {D.}~\bibnamefont
  {Bang}}, \bibinfo {author} {\bibfnamefont {T.}~\bibnamefont {Nozaki}},
  \bibinfo {author} {\bibfnamefont {D.~D.}\ \bibnamefont {Djayaprawira}},
  \bibinfo {author} {\bibfnamefont {M.}~\bibnamefont {Shiraishi}}, \bibinfo
  {author} {\bibfnamefont {Y.}~\bibnamefont {Suzuki}}, \bibinfo {author}
  {\bibfnamefont {A.}~\bibnamefont {Fukushima}}, \bibinfo {author}
  {\bibfnamefont {H.}~\bibnamefont {Kubota}}, \bibinfo {author} {\bibfnamefont
  {T.}~\bibnamefont {Nagahama}}, \bibinfo {author} {\bibfnamefont
  {S.}~\bibnamefont {Yuasa}}, \bibinfo {author} {\bibfnamefont
  {H.}~\bibnamefont {Maehara}}, \bibinfo {author} {\bibfnamefont
  {K.}~\bibnamefont {Tsunekawa}}, \bibinfo {author} {\bibfnamefont
  {Y.}~\bibnamefont {Nagamine}}, \bibinfo {author} {\bibfnamefont
  {N.}~\bibnamefont {Watanabe}}, \ and\ \bibinfo {author} {\bibfnamefont
  {H.}~\bibnamefont {Itoh}},\ }\href {\doibase 10.1063/1.3063674} {\bibfield
  {journal} {\bibinfo  {journal} {J. Appl. Phys.}\ }\textbf {\bibinfo {volume}
  {105}},\ \bibinfo {pages} {07C924} (\bibinfo {year} {2009})}\BibitemShut
  {NoStop}%
\bibitem [{\citenamefont {Ma}\ \emph {et~al.}(2011)\citenamefont {Ma},
  \citenamefont {Wang}, \citenamefont {Wei}, \citenamefont {Liu}, \citenamefont
  {Zhang},\ and\ \citenamefont {Han}}]{Ma2011prb}%
  \BibitemOpen
  \bibfield  {author} {\bibinfo {author} {\bibfnamefont {Q.~L.}\ \bibnamefont
  {Ma}}, \bibinfo {author} {\bibfnamefont {S.~G.}\ \bibnamefont {Wang}},
  \bibinfo {author} {\bibfnamefont {H.~X.}\ \bibnamefont {Wei}}, \bibinfo
  {author} {\bibfnamefont {H.~F.}\ \bibnamefont {Liu}}, \bibinfo {author}
  {\bibfnamefont {X.-G.}\ \bibnamefont {Zhang}}, \ and\ \bibinfo {author}
  {\bibfnamefont {X.~F.}\ \bibnamefont {Han}},\ }\href {\doibase
  10.1103/PhysRevB.83.224430} {\bibfield  {journal} {\bibinfo  {journal} {Phys.
  Rev. B}\ }\textbf {\bibinfo {volume} {83}},\ \bibinfo {pages} {224430}
  (\bibinfo {year} {2011})}\BibitemShut {NoStop}%
\bibitem [{\citenamefont {Drewello}\ \emph {et~al.}(2012)\citenamefont
  {Drewello}, \citenamefont {Ebke}, \citenamefont {Schafers}, \citenamefont
  {Kugler}, \citenamefont {Reiss},\ and\ \citenamefont
  {Thomas}}]{Drewello2012jap}%
  \BibitemOpen
  \bibfield  {author} {\bibinfo {author} {\bibfnamefont {V.}~\bibnamefont
  {Drewello}}, \bibinfo {author} {\bibfnamefont {D.}~\bibnamefont {Ebke}},
  \bibinfo {author} {\bibfnamefont {M.}~\bibnamefont {Schafers}}, \bibinfo
  {author} {\bibfnamefont {Z.}~\bibnamefont {Kugler}}, \bibinfo {author}
  {\bibfnamefont {G.}~\bibnamefont {Reiss}}, \ and\ \bibinfo {author}
  {\bibfnamefont {A.}~\bibnamefont {Thomas}},\ }\href {\doibase
  10.1063/1.3669913} {\bibfield  {journal} {\bibinfo  {journal} {J. Appl.
  Phys.}\ }\textbf {\bibinfo {volume} {111}},\ \bibinfo {pages} {07C701}
  (\bibinfo {year} {2012})}\BibitemShut {NoStop}%
\bibitem [{\citenamefont {Teixeira}\ \emph {et~al.}(2012)\citenamefont
  {Teixeira}, \citenamefont {Ventura}, \citenamefont
  {Fern\'{a}ndez-Garc\'{i}a}, \citenamefont {Araujo}, \citenamefont {Sousa},
  \citenamefont {Wisniowski},\ and\ \citenamefont {Freitas}}]{Teixeira2012apl}%
  \BibitemOpen
  \bibfield  {author} {\bibinfo {author} {\bibfnamefont {J.~M.}\ \bibnamefont
  {Teixeira}}, \bibinfo {author} {\bibfnamefont {J.}~\bibnamefont {Ventura}},
  \bibinfo {author} {\bibfnamefont {M.~P.}\ \bibnamefont
  {Fern\'{a}ndez-Garc\'{i}a}}, \bibinfo {author} {\bibfnamefont {J.~P.}\
  \bibnamefont {Araujo}}, \bibinfo {author} {\bibfnamefont {J.~B.}\
  \bibnamefont {Sousa}}, \bibinfo {author} {\bibfnamefont {P.}~\bibnamefont
  {Wisniowski}}, \ and\ \bibinfo {author} {\bibfnamefont {P.~P.}\ \bibnamefont
  {Freitas}},\ }\href {\doibase 10.1063/1.3687200} {\bibfield  {journal}
  {\bibinfo  {journal} {Appl. Phys. Lett.}\ }\textbf {\bibinfo {volume}
  {100}},\ \bibinfo {eid} {072406} (\bibinfo {year} {2012})}\BibitemShut
  {NoStop}%
\bibitem [{\citenamefont {Miao}\ \emph {et~al.}(2006)\citenamefont {Miao},
  \citenamefont {Chetry}, \citenamefont {Gupta}, \citenamefont {Butler},
  \citenamefont {Tsunekawa}, \citenamefont {Djayaprawira},\ and\ \citenamefont
  {Xiao}}]{Miao2006jap}%
  \BibitemOpen
  \bibfield  {author} {\bibinfo {author} {\bibfnamefont {G.-X.}\ \bibnamefont
  {Miao}}, \bibinfo {author} {\bibfnamefont {K.~B.}\ \bibnamefont {Chetry}},
  \bibinfo {author} {\bibfnamefont {A.}~\bibnamefont {Gupta}}, \bibinfo
  {author} {\bibfnamefont {W.~H.}\ \bibnamefont {Butler}}, \bibinfo {author}
  {\bibfnamefont {K.}~\bibnamefont {Tsunekawa}}, \bibinfo {author}
  {\bibfnamefont {D.}~\bibnamefont {Djayaprawira}}, \ and\ \bibinfo {author}
  {\bibfnamefont {G.}~\bibnamefont {Xiao}},\ }\href {\doibase
  10.1063/1.2162047} {\bibfield  {journal} {\bibinfo  {journal} {J. Appl.
  Phys.}\ }\textbf {\bibinfo {volume} {99}},\ \bibinfo {pages} {08T305}
  (\bibinfo {year} {2006})}\BibitemShut {NoStop}%
\bibitem [{\citenamefont {Drewello}\ \emph {et~al.}(2009)\citenamefont
  {Drewello}, \citenamefont {Sch\"{a}fers}, \citenamefont {Schebaum},
  \citenamefont {Khan}, \citenamefont {M\"{u}nchenberger}, \citenamefont
  {Schmalhorst}, \citenamefont {Reiss},\ and\ \citenamefont
  {Thomas}}]{Drewello2009prb}%
  \BibitemOpen
  \bibfield  {author} {\bibinfo {author} {\bibfnamefont {V.}~\bibnamefont
  {Drewello}}, \bibinfo {author} {\bibfnamefont {M.}~\bibnamefont
  {Sch\"{a}fers}}, \bibinfo {author} {\bibfnamefont {O.}~\bibnamefont
  {Schebaum}}, \bibinfo {author} {\bibfnamefont {A.~A.}\ \bibnamefont {Khan}},
  \bibinfo {author} {\bibfnamefont {J.}~\bibnamefont {M\"{u}nchenberger}},
  \bibinfo {author} {\bibfnamefont {J.}~\bibnamefont {Schmalhorst}}, \bibinfo
  {author} {\bibfnamefont {G.}~\bibnamefont {Reiss}}, \ and\ \bibinfo {author}
  {\bibfnamefont {A.}~\bibnamefont {Thomas}},\ }\href {\doibase
  10.1103/PhysRevB.79.174417} {\bibfield  {journal} {\bibinfo  {journal} {Phys.
  Rev. B}\ }\textbf {\bibinfo {volume} {79}},\ \bibinfo {pages} {174417}
  (\bibinfo {year} {2009})}\BibitemShut {NoStop}%
\bibitem [{\citenamefont {Moodera}\ \emph {et~al.}(1998)\citenamefont
  {Moodera}, \citenamefont {Nowak},\ and\ \citenamefont {van~de
  Veerdonk}}]{Moodera1998prl}%
  \BibitemOpen
  \bibfield  {author} {\bibinfo {author} {\bibfnamefont {J.~S.}\ \bibnamefont
  {Moodera}}, \bibinfo {author} {\bibfnamefont {J.}~\bibnamefont {Nowak}}, \
  and\ \bibinfo {author} {\bibfnamefont {R.~J.~M.}\ \bibnamefont {van~de
  Veerdonk}},\ }\href {\doibase 10.1103/PhysRevLett.80.2941} {\bibfield
  {journal} {\bibinfo  {journal} {Phys. Rev. Lett.}\ }\textbf {\bibinfo
  {volume} {80}},\ \bibinfo {pages} {2941} (\bibinfo {year}
  {1998})}\BibitemShut {NoStop}%
\bibitem [{\citenamefont {Wei}\ \emph {et~al.}(2010)\citenamefont {Wei},
  \citenamefont {Qin}, \citenamefont {Ma}, \citenamefont {Zhang},\ and\
  \citenamefont {Han}}]{Wei2010prb}%
  \BibitemOpen
  \bibfield  {author} {\bibinfo {author} {\bibfnamefont {H.-X.}\ \bibnamefont
  {Wei}}, \bibinfo {author} {\bibfnamefont {Q.-H.}\ \bibnamefont {Qin}},
  \bibinfo {author} {\bibfnamefont {Q.-L.}\ \bibnamefont {Ma}}, \bibinfo
  {author} {\bibfnamefont {X.-G.}\ \bibnamefont {Zhang}}, \ and\ \bibinfo
  {author} {\bibfnamefont {X.-F.}\ \bibnamefont {Han}},\ }\href {\doibase
  10.1103/PhysRevB.82.134436} {\bibfield  {journal} {\bibinfo  {journal} {Phys.
  Rev. B}\ }\textbf {\bibinfo {volume} {82}},\ \bibinfo {pages} {134436}
  (\bibinfo {year} {2010})}\BibitemShut {NoStop}%
\bibitem [{\citenamefont {Al'tshuler}\ and\ \citenamefont
  {Aronov}(1979)}]{Altshuler1979}%
  \BibitemOpen
  \bibfield  {author} {\bibinfo {author} {\bibfnamefont {B.~L.}\ \bibnamefont
  {Al'tshuler}}\ and\ \bibinfo {author} {\bibfnamefont {A.~G.}\ \bibnamefont
  {Aronov}},\ }\href@noop {} {\bibfield  {journal} {\bibinfo  {journal} {Sov.
  Phys. JETP}\ }\textbf {\bibinfo {volume} {50}},\ \bibinfo {pages} {968}
  (\bibinfo {year} {1979})}\BibitemShut {NoStop}%
\bibitem [{\citenamefont {Altshuler}\ \emph {et~al.}(1980)\citenamefont
  {Altshuler}, \citenamefont {Aronov},\ and\ \citenamefont
  {Lee}}]{Altshuler1980}%
  \BibitemOpen
  \bibfield  {author} {\bibinfo {author} {\bibfnamefont {B.~L.}\ \bibnamefont
  {Altshuler}}, \bibinfo {author} {\bibfnamefont {A.~G.}\ \bibnamefont
  {Aronov}}, \ and\ \bibinfo {author} {\bibfnamefont {P.~A.}\ \bibnamefont
  {Lee}},\ }\href {\doibase 10.1103/PhysRevLett.44.1288} {\bibfield  {journal}
  {\bibinfo  {journal} {Phys. Rev. Lett.}\ }\textbf {\bibinfo {volume} {44}},\
  \bibinfo {pages} {1288} (\bibinfo {year} {1980})}\BibitemShut {NoStop}%
\bibitem [{\citenamefont {Altshuler}\ and\ \citenamefont
  {Aronov}(1985)}]{Altshuler1985}%
  \BibitemOpen
  \bibfield  {author} {\bibinfo {author} {\bibfnamefont {B.~L.}\ \bibnamefont
  {Altshuler}}\ and\ \bibinfo {author} {\bibfnamefont {A.~G.}\ \bibnamefont
  {Aronov}},\ }in\ \href@noop {} {\emph {\bibinfo {booktitle}
  {Electron-Electron Interactions in Disordered Systems}}},\ \bibinfo {series}
  {Modern Problems in Condensed Matter Sciences}, Vol.~\bibinfo {volume} {10},\
  \bibinfo {editor} {edited by\ \bibinfo {editor} {\bibfnamefont
  {M.}~\bibnamefont {Pollak}}\ and\ \bibinfo {editor} {\bibfnamefont {A.~L.}\
  \bibnamefont {Efros}}}\ (\bibinfo  {publisher} {North-Holland},\ \bibinfo
  {address} {Amsterdam},\ \bibinfo {year} {1985})\ Chap.~\bibinfo {chapter}
  {1}, pp.\ \bibinfo {pages} {1--154}\BibitemShut {NoStop}%
\bibitem [{\citenamefont {Abrikosov}(1988)}]{Abrikosov1988}%
  \BibitemOpen
  \bibfield  {author} {\bibinfo {author} {\bibfnamefont {A.}~\bibnamefont
  {Abrikosov}},\ }\href@noop {} {\emph {\bibinfo {title} {Fundamentals of the
  theory of metals}}}\ (\bibinfo  {publisher} {Elsevier Science Pub. Co.},\
  \bibinfo {year} {1988})\BibitemShut {NoStop}%
\bibitem [{\citenamefont {Gershenzon}\ \emph {et~al.}(1986)\citenamefont
  {Gershenzon}, \citenamefont {Gubankov},\ and\ \citenamefont
  {Fale{\v{i}}}}]{Gershenzon1986jetp}%
  \BibitemOpen
  \bibfield  {author} {\bibinfo {author} {\bibfnamefont {M.~E.}\ \bibnamefont
  {Gershenzon}}, \bibinfo {author} {\bibfnamefont {V.~N.}\ \bibnamefont
  {Gubankov}}, \ and\ \bibinfo {author} {\bibfnamefont {M.~I.}\ \bibnamefont
  {Fale{\v{i}}}},\ }\href@noop {} {\bibfield  {journal} {\bibinfo  {journal}
  {Sov. Phys. JETP}\ }\textbf {\bibinfo {volume} {63}},\ \bibinfo {pages}
  {1287} (\bibinfo {year} {1986})}\BibitemShut {NoStop}%
\bibitem [{\citenamefont {Abrikosov}(2000)}]{Abrikosov2000prb}%
  \BibitemOpen
  \bibfield  {author} {\bibinfo {author} {\bibfnamefont {A.~A.}\ \bibnamefont
  {Abrikosov}},\ }\href {\doibase 10.1103/PhysRevB.61.7770} {\bibfield
  {journal} {\bibinfo  {journal} {Phys. Rev. B}\ }\textbf {\bibinfo {volume}
  {61}},\ \bibinfo {pages} {7770} (\bibinfo {year} {2000})}\BibitemShut
  {NoStop}%
\bibitem [{\citenamefont {Mazur}\ \emph {et~al.}(2007)\citenamefont {Mazur},
  \citenamefont {Gray}, \citenamefont {Zasadzinski}, \citenamefont {Ozyuzer},
  \citenamefont {Beloborodov}, \citenamefont {Zheng},\ and\ \citenamefont
  {Mitchell}}]{Mazur2007prb}%
  \BibitemOpen
  \bibfield  {author} {\bibinfo {author} {\bibfnamefont {D.}~\bibnamefont
  {Mazur}}, \bibinfo {author} {\bibfnamefont {K.~E.}\ \bibnamefont {Gray}},
  \bibinfo {author} {\bibfnamefont {J.~F.}\ \bibnamefont {Zasadzinski}},
  \bibinfo {author} {\bibfnamefont {L.}~\bibnamefont {Ozyuzer}}, \bibinfo
  {author} {\bibfnamefont {I.~S.}\ \bibnamefont {Beloborodov}}, \bibinfo
  {author} {\bibfnamefont {H.}~\bibnamefont {Zheng}}, \ and\ \bibinfo {author}
  {\bibfnamefont {J.~F.}\ \bibnamefont {Mitchell}},\ }\href {\doibase
  10.1103/PhysRevB.76.193102} {\bibfield  {journal} {\bibinfo  {journal} {Phys.
  Rev. B}\ }\textbf {\bibinfo {volume} {76}},\ \bibinfo {pages} {193102}
  (\bibinfo {year} {2007})}\BibitemShut {NoStop}%
\bibitem [{\citenamefont {Holweg}\ \emph {et~al.}(1992)\citenamefont {Holweg},
  \citenamefont {Caro}, \citenamefont {Verbruggen},\ and\ \citenamefont
  {Radelaar}}]{Holweg1992prb}%
  \BibitemOpen
  \bibfield  {author} {\bibinfo {author} {\bibfnamefont {P.~A.~M.}\
  \bibnamefont {Holweg}}, \bibinfo {author} {\bibfnamefont {J.}~\bibnamefont
  {Caro}}, \bibinfo {author} {\bibfnamefont {A.~H.}\ \bibnamefont
  {Verbruggen}}, \ and\ \bibinfo {author} {\bibfnamefont {S.}~\bibnamefont
  {Radelaar}},\ }\href {\doibase 10.1103/PhysRevB.45.9311} {\bibfield
  {journal} {\bibinfo  {journal} {Phys. Rev. B}\ }\textbf {\bibinfo {volume}
  {45}},\ \bibinfo {pages} {9311} (\bibinfo {year} {1992})}\BibitemShut
  {NoStop}%
\bibitem [{\citenamefont {Wei}\ \emph {et~al.}(2006)\citenamefont {Wei},
  \citenamefont {Pereverzev},\ and\ \citenamefont {Gershenson}}]{Wei2006}%
  \BibitemOpen
  \bibfield  {author} {\bibinfo {author} {\bibfnamefont {J.}~\bibnamefont
  {Wei}}, \bibinfo {author} {\bibfnamefont {S.}~\bibnamefont {Pereverzev}}, \
  and\ \bibinfo {author} {\bibfnamefont {M.~E.}\ \bibnamefont {Gershenson}},\
  }\href@noop {} {\bibfield  {journal} {\bibinfo  {journal} {Phys. Rev. Lett.}\
  }\textbf {\bibinfo {volume} {96}},\ \bibinfo {eid} {086801} (\bibinfo {year}
  {2006})}\BibitemShut {NoStop}%
\bibitem [{\citenamefont {Simmons}(1963)}]{Simmons1963jap}%
  \BibitemOpen
  \bibfield  {author} {\bibinfo {author} {\bibfnamefont {J.~G.}\ \bibnamefont
  {Simmons}},\ }\href {\doibase 10.1063/1.1702682} {\bibfield  {journal}
  {\bibinfo  {journal} {J. Appl. Phys.}\ }\textbf {\bibinfo {volume} {34}},\
  \bibinfo {pages} {1793} (\bibinfo {year} {1963})}\BibitemShut {NoStop}%
\bibitem [{\citenamefont {White}\ \emph {et~al.}(1985)\citenamefont {White},
  \citenamefont {Dynes},\ and\ \citenamefont {Garno}}]{White1985prb}%
  \BibitemOpen
  \bibfield  {author} {\bibinfo {author} {\bibfnamefont {A.~E.}\ \bibnamefont
  {White}}, \bibinfo {author} {\bibfnamefont {R.~C.}\ \bibnamefont {Dynes}}, \
  and\ \bibinfo {author} {\bibfnamefont {J.~P.}\ \bibnamefont {Garno}},\ }\href
  {\doibase 10.1103/PhysRevB.31.1174} {\bibfield  {journal} {\bibinfo
  {journal} {Phys. Rev. B}\ }\textbf {\bibinfo {volume} {31}},\ \bibinfo
  {pages} {1174} (\bibinfo {year} {1985})}\BibitemShut {NoStop}%
\bibitem [{\citenamefont {Gershenzon}\ \emph {et~al.}(1985)\citenamefont
  {Gershenzon}, \citenamefont {Gubankov},\ and\ \citenamefont
  {Fale{\v{i}}}}]{Gershenzon1985jetpl}%
  \BibitemOpen
  \bibfield  {author} {\bibinfo {author} {\bibfnamefont {M.~E.}\ \bibnamefont
  {Gershenzon}}, \bibinfo {author} {\bibfnamefont {V.~N.}\ \bibnamefont
  {Gubankov}}, \ and\ \bibinfo {author} {\bibfnamefont {M.~I.}\ \bibnamefont
  {Fale{\v{i}}}},\ }\href@noop {} {\bibfield  {journal} {\bibinfo  {journal}
  {Sov. Phys. JETP Lett.}\ }\textbf {\bibinfo {volume} {41}},\ \bibinfo {pages}
  {534} (\bibinfo {year} {1985})}\BibitemShut {NoStop}%
\bibitem [{\citenamefont {Lambe}\ and\ \citenamefont
  {Jaklevic}(1968)}]{Lambe1968pr}%
  \BibitemOpen
  \bibfield  {author} {\bibinfo {author} {\bibfnamefont {J.}~\bibnamefont
  {Lambe}}\ and\ \bibinfo {author} {\bibfnamefont {R.~C.}\ \bibnamefont
  {Jaklevic}},\ }\href {\doibase 10.1103/PhysRev.165.821} {\bibfield  {journal}
  {\bibinfo  {journal} {Phys. Rev.}\ }\textbf {\bibinfo {volume} {165}},\
  \bibinfo {pages} {821} (\bibinfo {year} {1968})}\BibitemShut {NoStop}%
\bibitem [{\citenamefont {Ashcroft}\ and\ \citenamefont
  {Mermin}(1976)}]{Ashcroft1976}%
  \BibitemOpen
  \bibfield  {author} {\bibinfo {author} {\bibfnamefont {N.~W.}\ \bibnamefont
  {Ashcroft}}\ and\ \bibinfo {author} {\bibfnamefont {N.~D.}\ \bibnamefont
  {Mermin}},\ }\href@noop {} {\emph {\bibinfo {title} {Solid state physics}}}\
  (\bibinfo  {publisher} {Harcourt},\ \bibinfo {year} {1976})\BibitemShut
  {NoStop}%
\bibitem [{\citenamefont {Wang}\ \emph {et~al.}(2007)\citenamefont {Wang},
  \citenamefont {Zeng}, \citenamefont {Han}, \citenamefont {Zhang},
  \citenamefont {Sun},\ and\ \citenamefont {Zhang}}]{Wang2007prb}%
  \BibitemOpen
  \bibfield  {author} {\bibinfo {author} {\bibfnamefont {Y.}~\bibnamefont
  {Wang}}, \bibinfo {author} {\bibfnamefont {Z.~M.}\ \bibnamefont {Zeng}},
  \bibinfo {author} {\bibfnamefont {X.~F.}\ \bibnamefont {Han}}, \bibinfo
  {author} {\bibfnamefont {X.~G.}\ \bibnamefont {Zhang}}, \bibinfo {author}
  {\bibfnamefont {X.~C.}\ \bibnamefont {Sun}}, \ and\ \bibinfo {author}
  {\bibfnamefont {Z.}~\bibnamefont {Zhang}},\ }\href {\doibase
  10.1103/PhysRevB.75.214424} {\bibfield  {journal} {\bibinfo  {journal} {Phys.
  Rev. B}\ }\textbf {\bibinfo {volume} {75}},\ \bibinfo {pages} {214424}
  (\bibinfo {year} {2007})}\BibitemShut {NoStop}%
\bibitem [{\citenamefont {Ando}\ \emph {et~al.}(2005)\citenamefont {Ando},
  \citenamefont {Miyakoshi}, \citenamefont {Oogane}, \citenamefont {Miyazaki},
  \citenamefont {Kubota}, \citenamefont {Ando},\ and\ \citenamefont
  {Yuasa}}]{Ando2005apl}%
  \BibitemOpen
  \bibfield  {author} {\bibinfo {author} {\bibfnamefont {Y.}~\bibnamefont
  {Ando}}, \bibinfo {author} {\bibfnamefont {T.}~\bibnamefont {Miyakoshi}},
  \bibinfo {author} {\bibfnamefont {M.}~\bibnamefont {Oogane}}, \bibinfo
  {author} {\bibfnamefont {T.}~\bibnamefont {Miyazaki}}, \bibinfo {author}
  {\bibfnamefont {H.}~\bibnamefont {Kubota}}, \bibinfo {author} {\bibfnamefont
  {K.}~\bibnamefont {Ando}}, \ and\ \bibinfo {author} {\bibfnamefont
  {S.}~\bibnamefont {Yuasa}},\ }\href {\doibase 10.1063/1.2077861} {\bibfield
  {journal} {\bibinfo  {journal} {Appl. Phys. Lett.}\ }\textbf {\bibinfo
  {volume} {87}},\ \bibinfo {eid} {142502} (\bibinfo {year}
  {2005})}\BibitemShut {NoStop}%
\bibitem [{\citenamefont {Du}\ \emph {et~al.}(2010)\citenamefont {Du},
  \citenamefont {Wang}, \citenamefont {Ma}, \citenamefont {Wang}, \citenamefont
  {Ward}, \citenamefont {Zhang}, \citenamefont {Wang}, \citenamefont {Kohn},\
  and\ \citenamefont {Han}}]{Du2010prb}%
  \BibitemOpen
  \bibfield  {author} {\bibinfo {author} {\bibfnamefont {G.~X.}\ \bibnamefont
  {Du}}, \bibinfo {author} {\bibfnamefont {S.~G.}\ \bibnamefont {Wang}},
  \bibinfo {author} {\bibfnamefont {Q.~L.}\ \bibnamefont {Ma}}, \bibinfo
  {author} {\bibfnamefont {Y.}~\bibnamefont {Wang}}, \bibinfo {author}
  {\bibfnamefont {R.~C.~C.}\ \bibnamefont {Ward}}, \bibinfo {author}
  {\bibfnamefont {X.-G.}\ \bibnamefont {Zhang}}, \bibinfo {author}
  {\bibfnamefont {C.}~\bibnamefont {Wang}}, \bibinfo {author} {\bibfnamefont
  {A.}~\bibnamefont {Kohn}}, \ and\ \bibinfo {author} {\bibfnamefont {X.~F.}\
  \bibnamefont {Han}},\ }\href {\doibase 10.1103/PhysRevB.81.064438} {\bibfield
   {journal} {\bibinfo  {journal} {Phys. Rev. B}\ }\textbf {\bibinfo {volume}
  {81}},\ \bibinfo {pages} {064438} (\bibinfo {year} {2010})}\BibitemShut
  {NoStop}%
\bibitem [{\citenamefont {Nishioka}\ \emph {et~al.}(2008)\citenamefont
  {Nishioka}, \citenamefont {Matsumoto}, \citenamefont {Tomita}, \citenamefont
  {Nozaki}, \citenamefont {Suzuki}, \citenamefont {Itoh},\ and\ \citenamefont
  {Yuasa}}]{Nishioka2008apl}%
  \BibitemOpen
  \bibfield  {author} {\bibinfo {author} {\bibfnamefont {S.}~\bibnamefont
  {Nishioka}}, \bibinfo {author} {\bibfnamefont {R.}~\bibnamefont {Matsumoto}},
  \bibinfo {author} {\bibfnamefont {H.}~\bibnamefont {Tomita}}, \bibinfo
  {author} {\bibfnamefont {T.}~\bibnamefont {Nozaki}}, \bibinfo {author}
  {\bibfnamefont {Y.}~\bibnamefont {Suzuki}}, \bibinfo {author} {\bibfnamefont
  {H.}~\bibnamefont {Itoh}}, \ and\ \bibinfo {author} {\bibfnamefont
  {S.}~\bibnamefont {Yuasa}},\ }\href {\doibase 10.1063/1.2988647} {\bibfield
  {journal} {\bibinfo  {journal} {Appl. Phys. Lett.}\ }\textbf {\bibinfo
  {volume} {93}},\ \bibinfo {eid} {122511} (\bibinfo {year}
  {2008})}\BibitemShut {NoStop}%
\bibitem [{\citenamefont {Klein}\ \emph {et~al.}(1973)\citenamefont {Klein},
  \citenamefont {L\'eger}, \citenamefont {Belin}, \citenamefont
  {D\'efourneau},\ and\ \citenamefont {Sangster}}]{Klein1973prb}%
  \BibitemOpen
  \bibfield  {author} {\bibinfo {author} {\bibfnamefont {J.}~\bibnamefont
  {Klein}}, \bibinfo {author} {\bibfnamefont {A.}~\bibnamefont {L\'eger}},
  \bibinfo {author} {\bibfnamefont {M.}~\bibnamefont {Belin}}, \bibinfo
  {author} {\bibfnamefont {D.}~\bibnamefont {D\'efourneau}}, \ and\ \bibinfo
  {author} {\bibfnamefont {M.~J.~L.}\ \bibnamefont {Sangster}},\ }\href
  {\doibase 10.1103/PhysRevB.7.2336} {\bibfield  {journal} {\bibinfo  {journal}
  {Phys. Rev. B}\ }\textbf {\bibinfo {volume} {7}},\ \bibinfo {pages} {2336}
  (\bibinfo {year} {1973})}\BibitemShut {NoStop}%
\bibitem [{\citenamefont {Lauhon}\ and\ \citenamefont
  {Ho}(2001)}]{Lauhon2001rsi}%
  \BibitemOpen
  \bibfield  {author} {\bibinfo {author} {\bibfnamefont {L.~J.}\ \bibnamefont
  {Lauhon}}\ and\ \bibinfo {author} {\bibfnamefont {W.}~\bibnamefont {Ho}},\
  }\href {\doibase 10.1063/1.1327311} {\bibfield  {journal} {\bibinfo
  {journal} {Review of Scientific Instruments}\ }\textbf {\bibinfo {volume}
  {72}},\ \bibinfo {pages} {216} (\bibinfo {year} {2001})}\BibitemShut
  {NoStop}%
\bibitem [{\citenamefont {Reed}(2008)}]{Reed2008mt}%
  \BibitemOpen
  \bibfield  {author} {\bibinfo {author} {\bibfnamefont {M.~A.}\ \bibnamefont
  {Reed}},\ }\href {\doibase 10.1016/S1369-7021(08)70238-4} {\bibfield
  {journal} {\bibinfo  {journal} {Materials Today}\ }\textbf {\bibinfo {volume}
  {11}},\ \bibinfo {pages} {46} (\bibinfo {year} {2008})}\BibitemShut {NoStop}%
\bibitem [{\citenamefont {Pekola}\ \emph {et~al.}(1994)\citenamefont {Pekola},
  \citenamefont {Hirvi}, \citenamefont {Kauppinen},\ and\ \citenamefont
  {Paalanen}}]{Pekola1994prl}%
  \BibitemOpen
  \bibfield  {author} {\bibinfo {author} {\bibfnamefont {J.~P.}\ \bibnamefont
  {Pekola}}, \bibinfo {author} {\bibfnamefont {K.~P.}\ \bibnamefont {Hirvi}},
  \bibinfo {author} {\bibfnamefont {J.~P.}\ \bibnamefont {Kauppinen}}, \ and\
  \bibinfo {author} {\bibfnamefont {M.~A.}\ \bibnamefont {Paalanen}},\ }\href
  {\doibase 10.1103/PhysRevLett.73.2903} {\bibfield  {journal} {\bibinfo
  {journal} {Phys. Rev. Lett.}\ }\textbf {\bibinfo {volume} {73}},\ \bibinfo
  {pages} {2903} (\bibinfo {year} {1994})}\BibitemShut {NoStop}%
\bibitem [{\citenamefont {Koster}\ \emph {et~al.}(2012)\citenamefont {Koster},
  \citenamefont {Klein}, \citenamefont {Siemons}, \citenamefont {Rijnders},
  \citenamefont {Dodge}, \citenamefont {Eom}, \citenamefont {Blank},\ and\
  \citenamefont {Beasley}}]{Koster2012rmp}%
  \BibitemOpen
  \bibfield  {author} {\bibinfo {author} {\bibfnamefont {G.}~\bibnamefont
  {Koster}}, \bibinfo {author} {\bibfnamefont {L.}~\bibnamefont {Klein}},
  \bibinfo {author} {\bibfnamefont {W.}~\bibnamefont {Siemons}}, \bibinfo
  {author} {\bibfnamefont {G.}~\bibnamefont {Rijnders}}, \bibinfo {author}
  {\bibfnamefont {J.~S.}\ \bibnamefont {Dodge}}, \bibinfo {author}
  {\bibfnamefont {C.-B.}\ \bibnamefont {Eom}}, \bibinfo {author} {\bibfnamefont
  {D.~H.~A.}\ \bibnamefont {Blank}}, \ and\ \bibinfo {author} {\bibfnamefont
  {M.~R.}\ \bibnamefont {Beasley}},\ }\href {\doibase
  10.1103/RevModPhys.84.253} {\bibfield  {journal} {\bibinfo  {journal} {Rev.
  Mod. Phys.}\ }\textbf {\bibinfo {volume} {84}},\ \bibinfo {pages} {253}
  (\bibinfo {year} {2012})}\BibitemShut {NoStop}%
\bibitem [{\citenamefont {Bhattacharya}\ \emph {et~al.}(2008)\citenamefont
  {Bhattacharya}, \citenamefont {May}, \citenamefont {te~Velthuis},
  \citenamefont {Warusawithana}, \citenamefont {Zhai}, \citenamefont {Jiang},
  \citenamefont {Zuo}, \citenamefont {Fitzsimmons}, \citenamefont {Bader},\
  and\ \citenamefont {Eckstein}}]{Bhattacharya2008prl}%
  \BibitemOpen
  \bibfield  {author} {\bibinfo {author} {\bibfnamefont {A.}~\bibnamefont
  {Bhattacharya}}, \bibinfo {author} {\bibfnamefont {S.~J.}\ \bibnamefont
  {May}}, \bibinfo {author} {\bibfnamefont {S.~G.~E.}\ \bibnamefont
  {te~Velthuis}}, \bibinfo {author} {\bibfnamefont {M.}~\bibnamefont
  {Warusawithana}}, \bibinfo {author} {\bibfnamefont {X.}~\bibnamefont {Zhai}},
  \bibinfo {author} {\bibfnamefont {B.}~\bibnamefont {Jiang}}, \bibinfo
  {author} {\bibfnamefont {J.-M.}\ \bibnamefont {Zuo}}, \bibinfo {author}
  {\bibfnamefont {M.~R.}\ \bibnamefont {Fitzsimmons}}, \bibinfo {author}
  {\bibfnamefont {S.~D.}\ \bibnamefont {Bader}}, \ and\ \bibinfo {author}
  {\bibfnamefont {J.~N.}\ \bibnamefont {Eckstein}},\ }\href {\doibase
  10.1103/PhysRevLett.100.257203} {\bibfield  {journal} {\bibinfo  {journal}
  {Phys. Rev. Lett.}\ }\textbf {\bibinfo {volume} {100}},\ \bibinfo {pages}
  {257203} (\bibinfo {year} {2008})}\BibitemShut {NoStop}%
\bibitem [{\citenamefont {Lepetit}\ \emph {et~al.}(2012)\citenamefont
  {Lepetit}, \citenamefont {Mercey},\ and\ \citenamefont
  {Simon}}]{Lepetit2012prl}%
  \BibitemOpen
  \bibfield  {author} {\bibinfo {author} {\bibfnamefont {M.-B.}\ \bibnamefont
  {Lepetit}}, \bibinfo {author} {\bibfnamefont {B.}~\bibnamefont {Mercey}}, \
  and\ \bibinfo {author} {\bibfnamefont {C.}~\bibnamefont {Simon}},\ }\href
  {\doibase 10.1103/PhysRevLett.108.087202} {\bibfield  {journal} {\bibinfo
  {journal} {Phys. Rev. Lett.}\ }\textbf {\bibinfo {volume} {108}},\ \bibinfo
  {pages} {087202} (\bibinfo {year} {2012})}\BibitemShut {NoStop}%
\bibitem [{\citenamefont {{Herranz}}\ \emph {et~al.}(2003)\citenamefont
  {{Herranz}}, \citenamefont {{Mart{\'{\i}}nez}}, \citenamefont
  {{Fontcuberta}}, \citenamefont {{S{\'a}nchez}}, \citenamefont
  {{Garc{\'{\i}}a-Cuenca}}, \citenamefont {{Ferrater}},\ and\ \citenamefont
  {{Varela}}}]{Herranz2003jap}%
  \BibitemOpen
  \bibfield  {author} {\bibinfo {author} {\bibfnamefont {G.}~\bibnamefont
  {{Herranz}}}, \bibinfo {author} {\bibfnamefont {B.}~\bibnamefont
  {{Mart{\'{\i}}nez}}}, \bibinfo {author} {\bibfnamefont {J.}~\bibnamefont
  {{Fontcuberta}}}, \bibinfo {author} {\bibfnamefont {F.}~\bibnamefont
  {{S{\'a}nchez}}}, \bibinfo {author} {\bibfnamefont {M.~V.}\ \bibnamefont
  {{Garc{\'{\i}}a-Cuenca}}}, \bibinfo {author} {\bibfnamefont {C.}~\bibnamefont
  {{Ferrater}}}, \ and\ \bibinfo {author} {\bibfnamefont {M.}~\bibnamefont
  {{Varela}}},\ }\href {\doibase 10.1063/1.1555372} {\bibfield  {journal}
  {\bibinfo  {journal} {J. Appl. Phys.}\ }\textbf {\bibinfo {volume} {93}},\
  \bibinfo {pages} {8035} (\bibinfo {year} {2003})}\BibitemShut {NoStop}%
\bibitem [{\citenamefont {Sun}\ \emph {et~al.}(1998)\citenamefont {Sun},
  \citenamefont {Abraham}, \citenamefont {Roche},\ and\ \citenamefont
  {Parkin}}]{Sun1998apl}%
  \BibitemOpen
  \bibfield  {author} {\bibinfo {author} {\bibfnamefont {J.~Z.}\ \bibnamefont
  {Sun}}, \bibinfo {author} {\bibfnamefont {D.~W.}\ \bibnamefont {Abraham}},
  \bibinfo {author} {\bibfnamefont {K.}~\bibnamefont {Roche}}, \ and\ \bibinfo
  {author} {\bibfnamefont {S.~S.~P.}\ \bibnamefont {Parkin}},\ }\href {\doibase
  10.1063/1.122068} {\bibfield  {journal} {\bibinfo  {journal} {Appl. Phys.
  Lett.}\ }\textbf {\bibinfo {volume} {73}},\ \bibinfo {pages} {1008} (\bibinfo
  {year} {1998})}\BibitemShut {NoStop}%
\bibitem [{\citenamefont {{O'Donnell}}\ \emph {et~al.}(2000)\citenamefont
  {{O'Donnell}}, \citenamefont {{Andrus}}, \citenamefont {{Oh}}, \citenamefont
  {{Colla}},\ and\ \citenamefont {{Eckstein}}}]{ODonnell2000apl}%
  \BibitemOpen
  \bibfield  {author} {\bibinfo {author} {\bibfnamefont {J.}~\bibnamefont
  {{O'Donnell}}}, \bibinfo {author} {\bibfnamefont {A.~E.}\ \bibnamefont
  {{Andrus}}}, \bibinfo {author} {\bibfnamefont {S.}~\bibnamefont {{Oh}}},
  \bibinfo {author} {\bibfnamefont {E.~V.}\ \bibnamefont {{Colla}}}, \ and\
  \bibinfo {author} {\bibfnamefont {J.~N.}\ \bibnamefont {{Eckstein}}},\ }\href
  {\doibase 10.1063/1.126210} {\bibfield  {journal} {\bibinfo  {journal} {Appl.
  Phys. Lett.}\ }\textbf {\bibinfo {volume} {76}},\ \bibinfo {pages} {1914}
  (\bibinfo {year} {2000})}\BibitemShut {NoStop}%
\bibitem [{\citenamefont {Takahashi}\ \emph {et~al.}(2003)\citenamefont
  {Takahashi}, \citenamefont {Sawa}, \citenamefont {Ishii}, \citenamefont
  {Akoh}, \citenamefont {Kawasaki},\ and\ \citenamefont
  {Tokura}}]{Takahashi2003prb}%
  \BibitemOpen
  \bibfield  {author} {\bibinfo {author} {\bibfnamefont {K.~S.}\ \bibnamefont
  {Takahashi}}, \bibinfo {author} {\bibfnamefont {A.}~\bibnamefont {Sawa}},
  \bibinfo {author} {\bibfnamefont {Y.}~\bibnamefont {Ishii}}, \bibinfo
  {author} {\bibfnamefont {H.}~\bibnamefont {Akoh}}, \bibinfo {author}
  {\bibfnamefont {M.}~\bibnamefont {Kawasaki}}, \ and\ \bibinfo {author}
  {\bibfnamefont {Y.}~\bibnamefont {Tokura}},\ }\href {\doibase
  10.1103/PhysRevB.67.094413} {\bibfield  {journal} {\bibinfo  {journal} {Phys.
  Rev. B}\ }\textbf {\bibinfo {volume} {67}},\ \bibinfo {pages} {094413}
  (\bibinfo {year} {2003})}\BibitemShut {NoStop}%
\bibitem [{\citenamefont {Wolf}(2012)}]{Wolf2012}%
  \BibitemOpen
  \bibfield  {author} {\bibinfo {author} {\bibfnamefont {E.}~\bibnamefont
  {Wolf}},\ }\href@noop {} {\emph {\bibinfo {title} {Principles of Electron
  Tunneling Spectroscopy}}},\ \bibinfo {edition} {second edition}\ ed.\
  (\bibinfo  {publisher} {Oxford University Press},\ \bibinfo {year}
  {2012})\BibitemShut {NoStop}%
\end{thebibliography}

%

\end{document}